\documentclass[onecolumn]{IEEEtran}

\usepackage{graphicx}
\graphicspath{{Figs/}}
\usepackage{amsmath,amssymb}
\usepackage{subcaption}
\usepackage{xcolor}
\usepackage{hyperref}
\usepackage{float}
\usepackage{enumerate}
\usepackage{diagbox}
\usepackage{soul}
\usepackage{pstool}

\newtheorem{theorem}{Theorem}
\newtheorem{proposition}{Proposition}
\newtheorem{corollary}{Corollary}
\newtheorem{lemma}{Lemma}
\newtheorem{definition}{Definition}
\newenvironment{remark}{\textit{Remark: }}{}
\newtheorem{example}{Example}[section]

\makeatletter
\renewcommand*\env@matrix[1][*\c@MaxMatrixCols c]{%
	\hskip -\arraycolsep
	\let\@ifnextchar\new@ifnextchar
	\array{#1}}
\makeatother

\newcommand{\bk}{\mathbf{k}}

\newcommand{\bn}{\mathbf{n}}

\newcommand{\bs}{\mathbf{s}}

\newcommand{\bx}{\mathbf{x}}
\newcommand{\by}{\mathbf{y}}

\newcommand{\bG}{\mathbf{G}}

\newcommand{\bN}{\mathbf{N}}

\newcommand{\bR}{\mathbf{R}}

\newcommand{\bT}{\mathbf{T}}

\newcommand{\bZ}{\mathbf{Z}}

\newcommand{\calK}{\mathcal{K}}

\newcommand{\ZZ}{\mathbb{Z}}

\newcommand{\FF}{\mathbb{F}}

\begin{document}

\title{Error-correcting codes for low latency streaming over multiple link relay networks}
\author{Gustavo Kasper Facenda, Elad Domanovitz, Ashish Khisti, Wai-Tian Tan and John Apostolopoulos}

\maketitle
\begin{abstract}
    This paper investigates the performance of streaming codes in low-latency applications over a multi-link three-node relayed network. The source wishes to transmit a sequence of messages to the destination through a relay. Each message must be reconstructed after a fixed decoding delay. The special case with one link connecting each node has been studied by Fong et. al \cite{Silas2019}, and a multi-hop multi-link setting has been studied by Domanovitz et. al \cite{GuaranteedMultihop2021}. The topology with three nodes and multiple links is studied in this paper. Each link is subject to a different number of erasures due to different channel conditions. An information-theoretic upper bound is derived, and an achievable scheme is presented. The proposed scheme judiciously allocates rates for each link based on the concept of delay spectrum. The achievable scheme is compared to two baseline schemes and the scheme proposed in \cite{GuaranteedMultihop2021}. Experimental results show that this scheme achieves higher rates than the other schemes, and can achieve the upper bound even in non-trivial scenarios. The scheme is further extended to handle different propagation delays in each link, something not previously considered in the literature. Simulations over statistical channels show that the proposed scheme can outperform the simpler baseline under practical models.
\end{abstract}

\section{Introduction}\label{intro}

Low-latency communication is essential in many real-time applications such as online gaming, cloud gaming, video conferences, tactile internet and virtual reality. In these applications, data packets are generated sequentially and must be transmitted to a destination within a strict delay constraint. During this transmission, packets might be lost over the network due to network congestion, software or hardware issues, or poor channel conditions. When packets are lost, error propagation can occur and significantly degrade the quality of the application, thus suitable methods for error correction are necessary.

Traditionally, error correction is handled through automatic repeat request (ARQ) or forward error correction (FEC). In long distance communications or strict low latency settings, ARQ is inherently inferior, as retransmissions will likely result in the delay constraint being violated \cite{lin1984automatic,lockitt1975selective,weldon1982improved,comroe1984arq}. For that reason, FEC schemes are considered more appropriate candidates for these applications. Indeed the early success of Skype is attributed to judicious use of error correction coding for supporting low-latency audio streaming \cite{Huang2010Skype}. Unfortunately, most traditional FEC codes, such as LDPC codes~\cite{gallager1962low,mackay1996near}, are not viable in low latency applications, as they involve long blocklengths, which result in significant buffering and wait times, increasing latency. For that reason, a new family of codes, denoted streaming codes, has been studied in the recent literature in order to establish fundamental limits of reliable low-latency communication under a variety of packet-loss models. 

The first work on streaming codes \cite{martinian2004burst} studied a point-to-point network under a maximal burst erasure pattern. In \cite{leong2012erasure}, the authors have studied, separately, burst erasures and arbitrary erasures. In \cite{badr2013streaming}, the authors have extended the erasure pattern, allowing for both burst erasures and arbitrary erasures. In particular, it was shown that random linear codes \cite{ho2003randomized} are optimal if we are concerned only with correcting arbitrary erasures. Other works that have further studied various aspects of low-latency streaming codes include \cite{JoshiWornell2012, Karzand2017, badr2017layered, badr2017fec, Rashmi2018, krishnan2018rate, fong2019optimal, domanovitz2019explicit, KrishnanLowField2020}.

While most of the prior work on streaming codes has focused on developing and improving codes for point-to-point communications, a network topology that is of practical interest involves a relay node between source and destination, i.e., a three-node network. This topology is motivated by numerous applications in which a gateway server, with access to enough computational power to process data, connects two end nodes. Such applications include cloud services, where a nearby gateway connects the user to an intranet where the server is located, or user-to-user communications that go through a centralized server, such as video conferencing. Motivated by such considerations, streaming codes for such a setting were first introduced in \cite{Silas2019}, extended to a multi-hop network in \cite{domanovitz2020streaming}, to a multi-user network in \cite{facenda2021streaming} and to a multi-hop multi-path setting in \cite{GuaranteedMultihop2021}. A recent work \cite{AdaptiveRelay} has also shown that allowing the relay to adapt its coding strategy, as opposed to the non-adaptive scheme proposed in \cite{Silas2019}, allows for rate improvement. The multi-hop multi-path setting is also studied in \cite{Cohen2019, Cohen2021}, where random linear codes are used, but adapt using a feedback signal.

In \cite{Silas2019, domanovitz2020streaming, facenda2021streaming, GuaranteedMultihop2021, AdaptiveRelay}, it is shown that, even under an arbitrary erasure model, random linear codes are suboptimal whenever relaying is involved, and codes that decode different symbols with different delays are used. Furthermore, in \cite{facenda2021streaming}, it is shown that allowing the relay to jointly encode data arriving from different users allows for rate improvement over only encoding separate paths. 


In this paper, we focus on the multi-link (or multi-path) setting in a three-node network. This network is useful in modeling mobile applications that simultaneously use both WiFi and cellular links, heterogeneous networks~\cite{Trestian2018heterogeneous}, software-defined networking~\cite{Hansen2015} and smart cities backhaul~\cite{SaadatMultipath2018}. Taking the results in \cite{facenda2021streaming} into account, we develop a coding scheme that allows for the relay to jointly encode data arriving from different paths, which has not been studied in \cite{domanovitz2020streaming, Cohen2021}. More precisely, we develop a framework that allows us to separately design point-to-point, single-link codes for each link, but that, at the same time, allows ``mixing'' packets at the relay, similar to \cite{facenda2021streaming}, but with multiple links in both hops. This novelty allows us to, for a wide range of parameters, achieve an upper bound on the rate for non-adaptive codes, and outperform naive extensions from the single-link three-node network \cite{Silas2019} as well as the scheme proposed in \cite{GuaranteedMultihop2021}. This framework is built upon the concept of delay spectrum, where each symbol is recovered with a different delay, based on the number of erasures it is subject to. By assuring that every symbol is recovered by the deadline, we assure the packet is recovered, and this approach allows us to carefully match delays in each hop by choosing where to transmit each symbol. This notion of different symbols having different delays is similar to multiplexed coding \cite{Multiplex2018, Fong2019multiplex}, where different streams have different delay deadlines. Thus, in developing tools for this framework, we also contribute towards these independent applications.

An additional challenge that arises in the multi-link setting is that different links may have a different propagation delay. Handling these different delays is not considered in \cite{GuaranteedMultihop2021}, and \cite{Cohen2021} considers only the worst propagation delay in each hop, and assumes that each hop has an equal propagation delay. In our work, we show that our framework is highly versatile and can be easily adapted to consider different propagation delays in each link with minor modifications in our algorithm. This is due to the nature of our framework, which directly works on the delay spectrum, thus the propagation delay is easily handled as an additional, constant delay for symbols transmitted through that link.

Finally, although we design codes for an adversarial model with a maximal number of arbitrary erasures, we simulate the codes designed using our scheme under statistical models, and compare it to a baseline scheme with same rate and delay deadline. In the i.i.d. case, our scheme strictly outperforms the baseline, and for Gilbert-Elliott channel model, it still outperforms for a considerable range of parameters.

Below we summarize our contributions.

\subsection*{Main contributions}

We introduce a model for low-latency streaming over a multi-link relay network and provide the following contributions:
\begin{itemize}
    \item We present an information theoretic upper-bound on the non-adaptive capacity for our model.
    \item We prove the optimality, in terms of delay spectrum, of the point-to-point, single-link streaming codes presented in \cite{facenda2021streaming}, which are of independent interest in multiplexed settings and are used as a building block in relayed settings.
    \item We propose a flexible delay-spectrum-constrained framework that can be used to design codes for general multi-link settings with general propagation delays in each link.
    \item We propose an empirical optimization algorithm under our framework that is used to design a scheme denoted as ``optimized symbol-wise decode-and-forward'' (OSWDF). 
    \item We numerically compare the rates achieved by OSWDF to extensions of \cite{Silas2019} and to \cite{GuaranteedMultihop2021} and show that we achieve noticeable higher rates.
    \item We compare codes designed with OSWDF against a baseline scheme under statistical models and show that we are able to outperform the baseline.
\end{itemize}

\section{Preliminaries}

\begin{figure*}[h!]
\centering
\includegraphics[scale=.9]{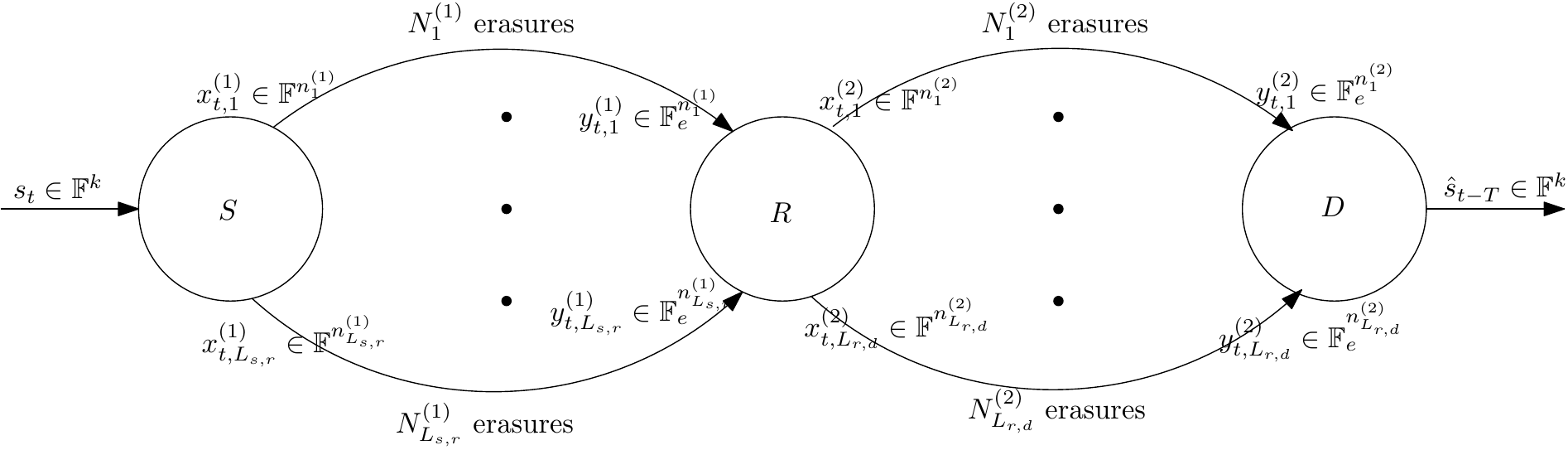}
\caption{Packets generated in the multi-link relay network at time $t$.}
\label{fig:multiLinkGeneral}
\end{figure*}

First, we introduce the basic setup of our problem. A source wishes to transmit a sequence of messages $\{\bs_t\}_{t=0}^{\infty}$ to a destination, with the help of a relay. There are $L_{s, r}$ links connecting the source to the relay, which we will denote as first hop, and $L_{r, d}$ links connecting the relay to destination, which we will denote as second hop. We assume that, on the discrete timeline, each link $i$ in the first hop introduces at most $N_i^{(1)}$ erasures, and each link $i$ in the second hop introduces at most $N_i^{(2)}$ erasures. The destination should decode the source packets with a maximum delay of $T$. For brevity, we denote by $\bN^{(1)} = [N_1^{(1)}, N_2^{(1)}, \ldots, N_{L_{s,r}}^{(1)}]$ and $\bN^{(2)} = [N_1^{(2)}, N_2^{(2)}, \ldots, N_{L_{r,d}}^{(2)}]$. For simplicity, we denote the set $\FF_e^n \triangleq \FF^n \cup \{*\}$. This setup is illustrated in Fig~\ref{fig:multiLinkGeneral} and formalized in this section.



\begin{definition}
	Let $\bn^{(1)} = [n^{(1)}_1, \ldots , n^{(1)}_{L_{s,r}}]$ and $\bn^{(2)} = [n^{(2)}_1, \ldots , n^{(2)}_{L_{r,d}}]$. An $(\bn^{(1)}, \bn^{(2)}, k, T)_{\FF}$-streaming code consists of the following:
	\begin{enumerate}
		\item A sequence of source messages $\{\bs_t\}_{t = 0}^{\infty}$, where $\bs_t \in \FF^{k}$. 
		\item A list of $L_{s, r}$ encoding functions $f^{(1)}_{t, i} : \underbrace{\FF^{k} \times \cdots \times \FF^{k} }_{t+1\textrm{ times}} \to \FF^{n_i^{(1)}}, \quad i \in \{1, \ldots, L_{s,r}\}$
		used by the source at time $t$ to generate $\bx_{t, i}^{(1)} = f^{(1)}_{t, i} (\bs_0, \bs_1, \ldots, \bs_t).$
		\item A list of $~L_{r, d}$ relaying functions $f^{(2)}_{t, i}: \underbrace{\FF_e^{n_1^{(1)}} \times \cdots \FF_e^{n_1^{(1)}}}_{t+1\textrm{ times}} \times  \cdots \times \underbrace{\FF_e^{n_{L_{s,r}}^{(1)}} \times \cdots \FF_e^{n_{L_{s,r}}^{(1)}}}_{t+1\textrm{ times}}, \quad i \in \{1, \ldots, L_{r, d}\}$
		used by the relay at time $t$ to generate $\bx_{t, i}^{(2)} = f^{(1)}_{t, i} (\{\by^{(1)}_{0, j}\}_{j=1}^{L_{s,r}}, \{\by^{(1)}_{1, j}\}_{j=1}^{L_{s,r}}, \ldots, \{\by^{(1)}_{t, j}\}_{j=1}^{L_{s,r}}), \quad i \in \{1, \ldots, L_{r, d}\} \label{eq:relayenc}$.
		\item A decoding function $\varphi_{t+T} = \underbrace{\FF_e^{n_1^{(2)}} \times \cdots \FF_e^{n_1^{(2)}}}_{t+T+1\textrm{ times}} \times  \cdots \times \underbrace{\FF_e^{n_{L_{r,d}}^{(2)}} \times \cdots \FF_e^{n_{L_{r,d}}^{(2)}}}_{t+T+1\textrm{ times}}$
		used by the destination at time $t + T$ to generate $\hat{\bs}_t = \varphi_{t+T}(\by^{(2)}_{0, j},\by^{(2)}_{1, j},\ldots,\by^{(2)}_{t+T, j} ), \quad j \in \{1, \ldots, L_{r, d}\}$.
	\end{enumerate}
\end{definition}
\begin{definition}
	An erasure sequence is a binary sequence denoted by $E_i^{(h)} \triangleq \{e^{(h)}_{t, i}\}_{t = 0}^{\infty}$, where $$e^{(h)}_{t,i}=\mathbf{1}\{\textrm{an erasure occurs at time $t$ in the $i$th link of the $h$th hop}\}.$$
	
	An $N$-erasure sequence is an erasure sequence $E_i^{(h)}$ that satisfies $\sum_{t=0}^{\infty} e^{(h)}_{t,i} = N$. In other words, an $N$-erasure sequence specifies $N$ arbitrary erasures on the discrete timeline. The set of $N$-erasure sequences is denoted by $\Omega_N$. 
\end{definition}
\begin{definition} \label{def:channel}
	The mapping $g_n : \FF^{n} \times \{0, 1\} \to \FF_e^{n}$ of an erasure channel is defined as
	\begin{align}
	g_n(\bx, e) = \begin{cases}
	\bx, &\textrm{~if~} e = 0\\
	*, &\textrm{~if~} e = 1  
	\end{cases} \label{eq:erasuremodel}
	\end{align}
	For any erasure sequences $E^{(1)}_i$ and any $(\bn^{(1)}, \bn^{(2)}, k, T)_{\FF}$-streaming code, the following input-output relation holds for the $i$th link in the first hop, for each $t \in \ZZ_+$:
	\begin{align}
	\by_{t, i}^{(1)} = g_{n_i^{(1)}}(\bx_{t, i}^{(1)}, e^{(1)}_{t, i}) \label{eq:erasurerelay}
	\end{align}
	where $E_i^{(1)} \in \Omega_{N_i^{(1)}}$, $i \in \{1, \ldots, L_{s,r}\}$.
	Similarly, the following input-output relation holds for the $i$th link the second hop, for each $t \in \ZZ_+$:
	\begin{align}
	\by_{t, i}^{(2)} = g_{n_i^{(2)}}(\bx_{t, i}^{(2)}, e^{(2)}_{t, i}) \label{eq:erasuredest}
	\end{align}
	where $E^{(2)}_{i} \in \Omega_{N_i^{(2)}}$, $i \in \{1, \ldots, L_{r,d}\}$.
\end{definition}

Recall that, at this moment, we are defining a zero-propagation delay setup, which models the scenario where all links in each hop have equal propagation delays. We extend this definition in Section~\ref{sec:propdelay}.

\begin{definition}
	An $(\bn^{(1)}, \bn^{(2)}, k, T)_{\FF}$-streaming code is said to be $(\bN^{(1)}, \bN^{(2)})$-achievable if, for any valid erasure sequences, for all $t \in \ZZ_+$ and all $\bs_t \in \FF^{k}$, we have $\hat{\bs}_t = \bs_t$.
\end{definition}
\begin{definition}
	The rate of an $(\bn^{(1)}, \bn^{(2)}, k, T)_{\FF}$-streaming code is $\frac{k}{ \max\left( \max( \bn^{(1)} ), \max(\bn^{(2)})   \right)   }$.
\end{definition}

\begin{remark}
    This definition of rate is an extension of the definition 
    of the rate of the three-node network \cite{Silas2019}. Further, the notion of capacity is natural as it considers the link with the maximum redundancy.
\end{remark}

\begin{definition}
	The $(T, \bN^{(1)}, \bN^{(2)})$-capacity, denoted by $C(T, \bN^{(1)}, \bN^{(2)})$, is the maximum achievable rate by $(\bn^{(1)}, \bn^{(2)}, k, T)_{\FF}$-streaming codes that are $(\bN^{(1)}, \bN^{(2)})$-achievable. 
\end{definition}

\subsection{Delay spectrum and decode-and-forward, key ingredients in the coding scheme}

In order to present our coding scheme, first let us define the notion of delay spectrum for a point-to-point, $L$-links code. This is a generalization of the definition of delay spectrum in \cite{Silas2019}, used to achieve capacity in the single-link topology.

\begin{definition}
	An $(\bn, k, \bT)_{\FF}$ point-to-point, $L$-links code, where $\bT = [T[1], \ldots, T[k]]$, consists of the following:
	\begin{enumerate}
		\item A sequence of source messages $\{\bs_t\}_{t = 0}^{\infty}$, where $\bs_t \in \FF^{k}$.
		\item A list of $L$ encoding functions
		\begin{align*}
		f_{t, i} &: \underbrace{\FF^{k} \times \cdots \times \FF^{k} }_{t+1\textrm{ times}} \to \FF^{n_i}, \quad i \in \{1, \ldots, L\}
		\end{align*}
		used by the transmitter at time $t$ to generate $$\bx_{t, i} = f_{t, i} (\bs_0, \bs_1, \ldots, \bs_t).$$
		\item A list of $k$ decoding function 
		\begin{align*}
		\varphi_{t+T[j]} &= \underbrace{\FF_e^{\bn_1} \times \cdots \FF_e^{\bn_1}}_{t+T[j]+1\textrm{ times}} \times \cdots \times \underbrace{\FF_e^{\bn_L} \times \cdots \FF_e^{\bn_L}}_{t+T[j]+1\textrm{ times}}, \quad j \in \{1, \ldots, k\}
		\end{align*}
		used by the receiver at time $t + T[j]$ to generate $\hat{\bs}_t[j]$.
	\end{enumerate}
\end{definition}
\begin{definition}
	An $(\bn, k, \bT)_{\FF}$ point-to-point, $L$-links code is said to achieve delay spectrum $\bT$ under $\bN$ erasures if, for any $E_i' \in \Omega_{n_i}$, 
	\begin{align*}
	&\varphi_{t + T[j]}( g_n(\bx_{0, i}, e_{0, i}'), \ldots, g_n(\bx_{t + T[j], i}, e'_{t + T[j], i})) = \bs_t[j], \quad i \in \{1, \ldots, L\}
	\end{align*}
\end{definition}

For the relaying strategy, let us now introduce the concept of symbol-wise decode-and-forward. In this strategy, the relaying function employed by the code first decodes the source packets transmitted by the source, and then encodes them again. Below, we formally define this strategy.

\begin{definition}\label{def:swdf}
	Assume the source node transmits its source messages $\{\bs_t\}_t^{\infty}$ to the relay using an $(\bn^{(1)}, k, \bT^{(1)})_{\FF}$ point-to-point, $L$-links code with decoding functions $\varphi^{(1)}_{t+ T^{(1)}[j]}$. Then, a relay is said to employ a symbol-wise decode-and-forward if the following holds:
	\begin{itemize}
		\item The relay employs the decoding functions $\varphi^{(1)}_{t + T^{(1)}[j]}$ at time $t + T^{(1)}[j]$ to estimate the $j$-th source symbol $\hat{\bs}_t[j]$.
		\item The relay employs an $(\bn^{(2)}, k, \bT^{(2)})_{\FF}$ point-to-point, $L$-links code to transmit a sequence of relay messages $\{ \bs'_t \}_{t = 0}^{\infty}$ to the destination.
		\item The relay messages are given by 
 		\begin{align}
 		\bs'_t[j'] = \hat{\bs}_{t - (T^{(1)}[j])}[j]. \label{eq:relaymessage}
 		\end{align}
		That is, the message used to encode in the relay is simply the relay's estimate of past messages transmitted by the source node.
	\end{itemize}
\end{definition}
\begin{remark}
    The operation performed in \eqref{eq:relaymessage} can be seen as a simple relabeling of the source packets. As long as we appropriately match the delay spectra of both hops, such relabeling can be helpful, as we now may treat each hop as a point-to-point hop, but with different delay constraints. This is exemplified in Appendix~\ref{ex:singlelink}.
\end{remark}

An example of how to achieve the upper bound for the single-link three-node network employing these strategies is presented in the Appendix \ref{ex:singlelink} as Example~1.

For the remaining of the paper, we use $\bT^{(1)}$ to denote the delay spectrum of the code used in the first hop and $\bT^{(2)}$ to denote the delay spectrum of the code used in the second hop.

\begin{lemma}\label{lemma:df}
	Assume there exists an $(\bn^{(1)}, k, \bT^{(1)})_{\FF}$ point-to-point, $L$-links code that achieves the delay spectrum $\bT^{(1)}$ under $\bN^{(1)}$ erasures and an $(\bn^{(2)}, k, \bT^{(2)})_{\FF}$ point-to-point, $L$-links code that achieves the delay spectrum $\bT^{(2)}$ under $\bN^{(2)}$ erasures. Then, there exists a streaming code that achieves the delay spectrum
	\begin{align}
	\bT = \bT^{(1)} + \pi \bT^{(2)}
	\end{align}
	under $\bN^{(1)}$ erasures in the first hop and $\bN^{(2)}$ erasures in the second hop, where $\pi$ is any permutation matrix.
\end{lemma}

\subsection{Delay spectrum of single-link codes and useful operations}
In the multiple link setting, it is interesting to consider concatenations of codes. As will be seen in the following section, this is an important idea in order to achieve higher rates. 

\begin{definition}
	A concatenation of an $(\bn', k', \bT')_{\FF}$ point-to-point, $L$-links code with an $(\bn'', k'', \bT'')_{\FF}$ point-to-point, $L$-links is an $(\bn' + \bn'', k' + k'', [\bT', \bT''])$ point-to-point, $L$-links code with the following properties
	\begin{itemize}
		\item Let $\{f'_{t, i}\}_{i = 1}^{L}$ be the list of encoding functions of the first code and $\{f''_{t, i}\}_{i = 1}^{L}$ be the list of encoding functions of the second code. The list of encoding functions of the concatenated code is given by $\{[f'_{t, i}, f''_{t, i}]\}_{i=1}^{L}$, where $[x, y]$ denotes the concatenation of a vector $x$ and a vector $y$.
		\item Let $\{\varphi'_{t + T'[j]}\}_{j=1}^{k'}$ be the list of decoding functions of the first code and $\{\varphi''_{t + T''[j]}\}_{j=1}^{k''}$ be the list of decoding functions of the second code. The list of decoding functions of the concatenated code is given by $\{\varphi'_{t + T'[j]}\}_{j=1}^{k'} \frown \{\varphi''_{t + T''[j]}\}_{j=1}^{k''}$, where $x \frown y$ denotes the concatenation of two lists $x$ and $y$.
	\end{itemize}
	\label{def:concatenation}
\end{definition}
	


\begin{lemma}\label{lemma:conc}
	If there exists an $(\bn', k', \bT')_{\FF}$ point-to-point $L$-links code that achieves delay spectrum $\bT'$ under $\bN$ erasures, and an $(\bn'', k'', \bT'')_{\FF}$ point-to-point $L$-links code that achieves delay spectrum $\bT''$ under $\bN$ erasures, then there exists an $(\bn' + \bn'', k' + k'', [\bT', \bT''])_{\FF}$ point-to-point $L$-links code that achieves delay spectrum $[\bT', \bT'']$ under $\bN$ erasures.  
\end{lemma}

Another useful operation that can be made is simply permuting the source symbols.
\begin{lemma}\label{lemma:permutation}
	Assume a delay spectrum $\bT$ is achievable under $N$ erasures by some code. Then, any permutation $\pi \bT$, where $\pi$ is a permutation matrix, of this delay spectrum is also achievable under $N$ erasures.
\end{lemma}

Since any permutation of an achievable delay spectrum is also achievable, we may instead describe the delay spectrum of a code by stating how many symbols are transmitted with some delay. For example, consider the code in Table~\ref{tab:exampcode}. Instead of saying it achieves a delay spectrum of $[2, 1, 3, 2, 1]$ under one erasure, we could describe it as having two symbols with delay 1, two symbols with delay 2 and one symbol with delay 3. Both of these descriptions are interchangeable, and both may be useful depending on the analysis being done. We call this equally-delayed symbols grouping.


\begin{definition}
	Consider a delay spectrum $\bT = [T[1], T[2], \ldots, T[k]]$. An equally-delayed symbols grouping description of such delay spectrum is given by a list of tuples
	$$\bG = [(T^{(g)}[1], k^{(g)}[1]),\ldots,(T^{(g)}[\ell^{(g)}], k^{(g)}[\ell^{(g)}])]$$ where $\ell^{(g)}$ is the length of the list. For simplicity, we assume $T^{(g)}[1] \geq T^{(g)}[2] \geq \cdots \geq T^{(g)}[\ell^{(g)}]$, therefore, $T^{(g)}[1] = \max(\bT)$ and $T^{(g)}[\ell^{(g)}] = \min(\bT)$.  Furthermore, we define
	\begin{itemize}
		\item $\bT^{(g)} = [T^{(g)}[1], \ldots, T^{(g)}[\ell^{(g)}]]$ as the ordered list of possible delays.  
		\item $\bk^{(g)} = [k^{(g)}[1], \ldots, k^{(g)}[\ell^{(g)}]]$, where $\sum_{i = 1}^{\ell^{(g)}} k^{(g)}[i] = k$ as the ordered list of number of symbols associated with each delay. 
	\end{itemize}
\end{definition}

Under this description, the number of symbols with some delay $\tau$ in a concatenated code is simply the sum of the number of symbols with that delay in each code that was concatenated. A detailed example of the concatenation operation and this notation is given in Example~\ref{ex:concatenation}.

Before we present the two main results of this section, let us make a distinction between a systematic code and a ``systematic code with respect to its own message''. In streaming codes literature, a systematic code transmits the source packets $\bs$ as information symbols \cite{badr2017layered}. In our definition, a ``systematic code with respect to its own message'' is systematic with respect to its own source sequence, i.e., $\bs$ for the source node and $\bs'$ as defined in \eqref{eq:relaymessage} for the relay node. This distinction is formalized below.
\begin{definition}
	An $(n, k, \bT)_{\FF}$ point-to-point, single-link code is said to be systematic with respect to its own sequence of source messages if $\bx_{t}[j] = \bs_t[j], j \in \{1, \ldots, k\}$
	for all $t$. 
\end{definition}
For the remaining of the paper, when referring to point-to-point codes, we omit the ``with respect to its own sequence of source messages'', and simply call such codes systematic.

In Example~\ref{ex:singlelink}, we have used diagonal interleaving MDS codes which helped us achieve the upper bound. Examples of diagonal interleaving MDS codes are given in the tables in the example. The following Lemma is proved in \cite{Silas2019}
\begin{lemma}\label{lemma:intMDS}
	A systematic $(N+k, k, \bT)_{\FF}$ point-to-point, single link diagonal interleaving MDS code achieves the delay spectrum $\bT = [N, \ldots, N + k - 1]$ under $N$ erasures.
\end{lemma}


We now state one main result of our paper: a lower bound on the achievable delay spectrum for a single-link point-to-point code.

\begin{lemma}[Lower Bound] \label{lemma:lb}
	For a systematic point-to-point $(n, k)_{\FF}$ single-link code subject to $N$ erasures, the delay of the $j$-th equally-delayed group is lower bounded by
	\begin{align}
	\bT^{(g)}[j] &\geq \frac{N n}{n - k}\left(1 - \sum_{\ell = 1}^{j-1} \frac{k^{(g)}[\ell]}{n} \right) - 1. \label{eq:DelayConLowBound}
	\end{align}
\end{lemma}

\begin{corollary}\label{corollary:bounds}
	Assume $\bT^{(g)} = [T^{(g)}[1], T^{(g)}[1] - 1, \ldots, N]$, i.e., the possible delays are separated by one, and the minimum delay is $N$. Then, the following holds for any $(n, k, \bT)_{\FF}$ point-to-point single-link code under $N$ erasures
	\begin{align*}
	k^{(g)}[1] \geq n - \frac{T^{(g)}[1]}{N}(n-k), \quad \textrm{and} \quad k^{(g)}[\ell^{(g)}] \leq \frac{n - k}{N} 
	\end{align*}
\end{corollary}

The implication of this corollary is that there is a minimum number of symbols that need to be transmitted at the worst possible delay (i.e., $T^{(g)}[1]$), and a maximum number of symbols that can be transmitted at the smallest possible delay.

\begin{corollary}\label{corollary:recursion}
    Assume $\bT^{(g)} = [T^{(g)}[1], T^{(g)}[1] - 1, \ldots, N]$, i.e., the possible delays are separated by one, and the minimum delay is $N$. Then, the optimal\footnote{At this point, optimal means ``the best we can hope for'', not necessarily achievable. We nonetheless use the word optimal as Lemma~\ref{lemma:achiev} proves that it is, indeed, achievable.} number of symbols transmitted with each delay under $N$ erasures, for any $(n, k, \bT)_{\FF}$ point-to-point single-link code is given by
    \begin{align}
    k^{(g)}[1] = n - \frac{T^{(g)}[1]}{N}(n-k), \quad \textrm{and} \quad k^{(g)}[j] = \frac{n - k}{N}, \forall j \in \{2, 3, \ldots, \ell^{(g)}\}  \label{eq:bestsymbols}
    \end{align}
\end{corollary}
\begin{remark}
    The results from Corollary~\ref{corollary:recursion} match nicely with the ones from Corollary~\ref{corollary:bounds}: it is optimal to transmit the most symbols at the lowest delay, and the fewest in the highest. 
\end{remark}

\begin{remark}
	Note that $k^{(g)}[1] \leq k^{(g)}[j]$ for all $j \geq 2$ since
	\begin{align*}
	n - \frac{T^{(g)}[1]}{N}(n - k) \leq \frac{1}{N}(n - k) \Longrightarrow 
	T^{(g)}[1](n - k) \geq Nn - (n - k) \Longrightarrow 
	T^{(g)}[1] \geq \frac{N n}{n - k} - 1
	\end{align*}
	which is the result from Lemma~\ref{lemma:lb}. 
\end{remark}



Another useful question to be answered in designing the codes is, assuming there is a constraint on the number of symbols to be transmitted in each delay slot, how many symbols can be transmitted? This is answered in Corollary~\ref{corollary:maxsym} below.

\begin{corollary}\label{corollary:maxsym}
	Assume there is a maximum number of symbols we are allowed to transmit at each delay, that is, $\bk^{(g)} \leq \bk^{\textrm{con}}$ is a constraint. Then, the maximum number of symbols $k$ that can be transmitted by an $(n, k, \bT)_{\FF}$ point-to-point single-link code while still achieving the desired delay spectrum under $N$ erasures is
	\begin{align}
	k \leq n - n \cdot N \cdot \frac{\left( 1 - \sum_{\ell = 1}^{j-1} \frac{k^{\textrm{con}}[\ell]}{n}  \right)}{T^{(g)}[j] + 1} \forall j \in \{1, 2, \ldots, \ell^{(g)}\}
	\end{align}
\end{corollary}

\begin{remark}
    Since the bound only depends on $\sum_{\ell=1}^{j-1} k^{\textrm{con}}[\ell]$, we can relax the constraint to $\sum_{\ell=1}^{j} k^{(g)}[\ell] \leq \sum_{\ell=1}^{j} k^{\textrm{con}}[\ell]$ for all $j$ and still get the same bound. Proof follows exactly as above.
\end{remark}

This completes the statements of the fundamental limits using point-to-point single-link codes. In the sequence, we state the achievable result previously proven in \cite{facenda2021streaming}. 

\begin{lemma}[Achievability] \label{lemma:achiev}
	Let \begin{align*}
	\bT^{(g)} = [T^{(g)}[1], T^{(g)}[2], \ldots, T^{(g)}[\ell^{(g)}]] = [T^{(g)}[1], T^{(g)}[1] - 1, \ldots, N+1, N]
	\end{align*}
	and $\frac{n-k}{N}$ be an integer.
	Then, there exists an $(n, k, \bT)_{\FF}$ point-to-point single-link code that can transmit $k^{(g)}[1] = n - \frac{T^{(g)}[1]}{N}(n-k)$ symbols with delay $T^{(g)}[1]$ and $k^{(g)}[j] = \frac{n - k}{N}$ for all other delays.
\end{lemma}

This implies the bound given in Corollary~\ref{corollary:recursion} is achievable, which is an equivalent statement to Lemma~\ref{lemma:lb}. 

\begin{remark}
    While this code construction is done as a tool for us to build upon in the three-node network, it also is of independent interest and could be used in applications which require different data to be recovered at different delay constraints by the destination, such as QUIC and video/audio conferencing.
\end{remark}

\subsection{Separate coding strategy} \label{sec:sepcode}

For the purposes of further development of our coding scheme, it suffices to use separate coding in each link, and not jointly encode and decode across all links. This is useful for code design, analysis and implementation complexity of both the encoder and decoder. Below, we formally define a separate coding strategy.


\begin{definition}
	An $(\bn, k, \bT)_{\FF}$ point-to-point, $L$-links code is said to employ a separate coding strategy if its encoding functions have the form
	\begin{align*}
	f_{t, i} &: \underbrace{\FF^{k_i} \times \cdots \times \FF^{k_i} }_{t+1\textrm{ times}} \to \FF^{n_i}, \quad i \in \{1, \ldots, L\}
	\end{align*}
	used by the transmitter at time $t$ to generate
	\begin{align*}
	\bx_{t, i} &= f_{t, i}(\bs_0[ j \in \calK_i ], \ldots, \bs_t[ j \in \calK_i])
	, \textrm{~where~} |\calK_i| = k_i \\
	\calK_i \cap \calK_{i'} &= \emptyset, i \neq i'
	, \quad |\cup_{i = 1}^{L} \calK_i| = k
	\end{align*}
	That is, each one of the $L$ functions depends on different symbols of the source symbols. 
\end{definition}

W.l.o.g., we may assume $\calK_1 = \{1, \ldots, k_1\}$, $\calK_2 = \{k_1+1, \ldots, k_1 + k_2\}$, and so on. 

\begin{lemma} \label{lemma:sepcod}
	Let $\{\bT_i\}_{i = 1}^{L}$ be a list of delay spectrum of $L$ single-link codes. If each $i$th delay spectrum $\bT_i$ is achievable under $N_i$ erasures, then the concatenation of all the delay spectra is achievable under $\bN = [N_1, N_2, \ldots, N_L]$.
\end{lemma}

Similar to concatenating codes, under the equally-delayed symbol description, the delay spectrum of the $L$-links code constructed with separate coding strategy can be found by appropriately summing the number of symbols with each delay from each single-link code.

Also, recall that any permutation of an achievable delay spectrum is also achievable. In the relay setting, what this implies is that symbols that have been transmitted through one link in the first hop do not need to be transmitted through the same link in the second hop. This is another important idea that allows us to achieve a better performance in the multiple-link scenario.

Employing both the decode-and-forward strategy described earlier and the separate coding strategy described in this section is useful, both for code construction and analysis. 

Before we continue, let us recall and introduce some notation. For the remaining of the paper, $\bT^{(h)}_i$ denotes the delay spectrum of the code used in the $i$th link in the $h$th hop, and $\bT^{(h)}$ without the subscript denotes the delay spectrum of the overall point-to-point code in the $h$th hop, which, under separate coding, consists of the concatenation and possible permutation of all $\{\bT^{(h)}_i\}$, with $i$ from 1 to the number of links in the $h$th hop. We denote by $k^{(h)}_i$ the number of source symbols transmitted in the $i$th link in the $h$th hop, and $n^{(h)}_i$ the number of channel symbols used by the code to transmit those source symbols. For brevity, some times we refer to the vectors $\bk^{(h)} = [k^{(h)}_1,k^{(h)}_2,\ldots,k^{(h)}_L]$ and $\bn^{(h)} = [n^{(h)}_1,n^{(h)}_2,\ldots,n^{(h)}_L]$ where $L$ is the number of links in the $h$th hop. Similarly, we define $R^{(h)}_i = \frac{k^{(h)}_i}{n^{(h)}_i}$ and $\bR^{(h)} = [R^{(h)}_1,R^{(h)}_2,\ldots,R^{(h)}_L] $. Finally, we denote by $k^{(h)} = \sum_{i=1}^{L_{s,r}} k^{(h)}_i$.

\subsection{Motivating example}
\begin{figure}[h!]
\centering
\includegraphics[scale=.95]{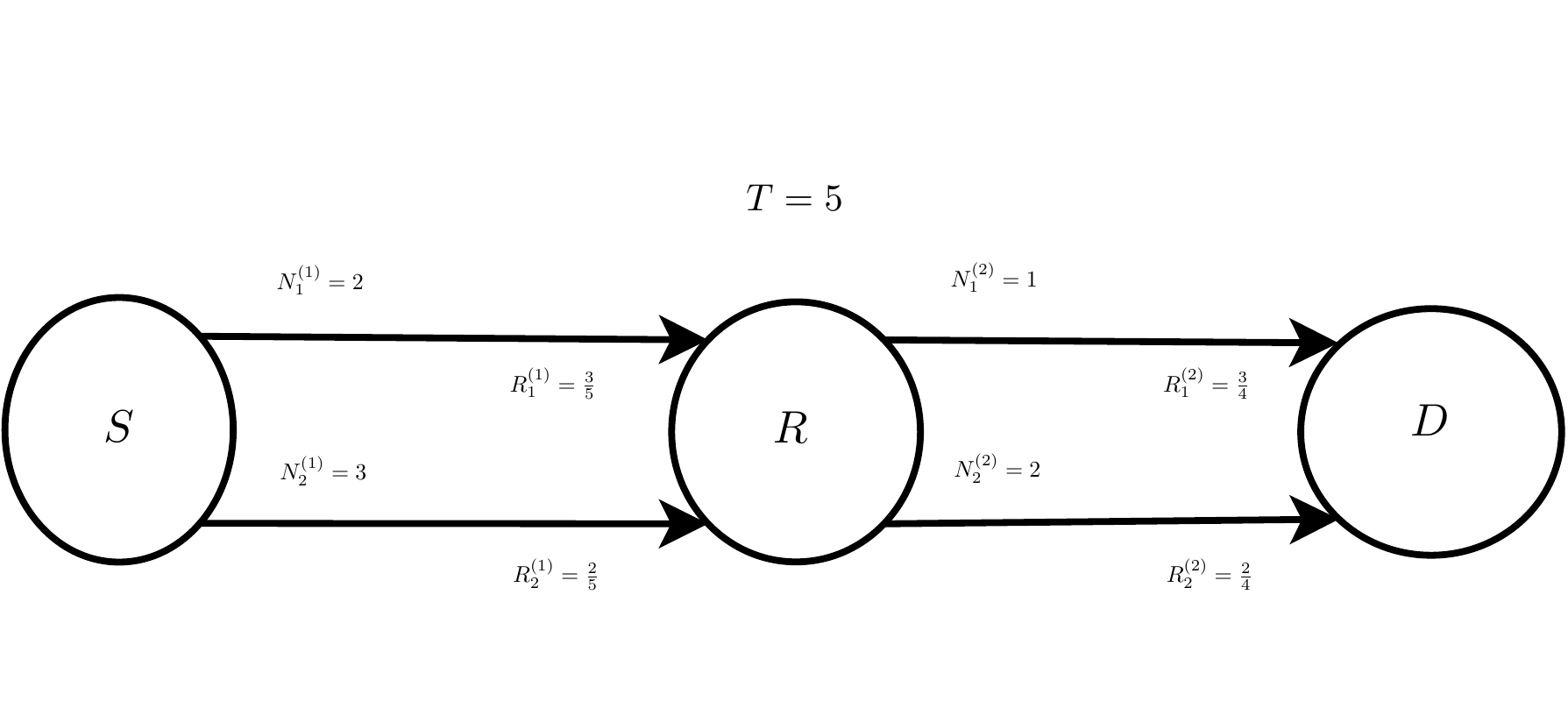}
\caption{Example of a setting with different number of erasures..}
\label{fig:achievex}
\end{figure}

As a motivation for the use of codes employing the strategies we mentioned previously, let us consider the example setting in Fig.~\ref{fig:achievex}. To the best of our knowledge, no prior work has studied a similar setting, and designing codes for it is a challenging open problem.


We propose a decode-and-forward separate coding code with parameters $k_1^{(1)} = 12$, $k_2^{(1)} = 8$, $k^{(2)}_1 = 11$ and $k^{(2)}_2 = 9$. The delay spectra for each link and the way to achieve it is summarized below
\begin{itemize}
	\item $\bT^{(1)}_1 = [2,2,2,2,3,3,3,3,4,4,4,4]$. Achieved by using four concatenated $(5, 3)$ codes. We use $n^{(1)}_1 = 20$.
	\item $\bT^{(1)}_2 = [3, 3, 3, 3, 4, 4, 4, 4]$. Achieved by using four concatenated $(5, 2)$ codes. We use $n^{(1)}_2 = 20$.
	\item $\bT^{(2)}_1 = [1, 1, 1, 1,1,1,1,1,1,2,2]$. Achieved by concatenating seven $(2, 1)$ codes and two $(3, 2)$ codes. We use $n^{(2)}_1 = 20$.
	\item $\bT^{(2)}_2 = [2, 2, 2, 2, 2, 3, 3, 3, 3]$. Achieved by concatenating four $(4, 2)$ codes and one $(3, 1)$ code. We use $n^{(2)}_2 = 19$.
\end{itemize}

Then, by appropriately matching the delay spectra of each hop, similar to what is done in Example~\ref{ex:singlelink}, we are able to recover all source packets by time $T = 5$ in the destination. To see that, note that there are $8$ symbols with delay $4$ in the first hop, which can be paired with $8$ symbols with delay $1$; there are $8$ symbols with delay $3$, which can be paired with seven symbols with delay $2$ and one remaining symbol with delay $1$; and $4$ symbols with delay $2$, which can be paired with the remaining four symbols with delay $3$. Therefore, this is an achievable code for the presented setting and its rate is 1. As we will show in Section~\ref{sec:upbound}, this is the upper bound, and thus the code is capacity-achieving.

Further, as we will show in Section~\ref{sec:baseline}, the MWDF can only achieve a rate of 3/4, and the natural extension of \cite{Silas2019}, which we call CSWDF, can only achieve a rate of 8/9.

\section{Upper Bound} \label{sec:upbound}

Similar to \cite{Silas2019}, we propose an upper bound on the capacity. 
The result is a direct extension of the previous result, in which we upper bound the rate of each hop by analyzing a specific erasure pattern in the other hop which represents an improved channel compared to the actual model. 
More precisely, when analyzing the upper bound on the rate for the first hop $R^{(1)}$, we assume a burst of $N^{(2)}_{\textrm{min}} = \min(\bN^{(2)})$ erasures in the second hop, and when analyzing the upper bound on the rate for the second hop $R^{(2)}$, we assume a burst of $N^{(1)}_{\textrm{min}} = \min(\bN^{(1)})$ erasures in the first hop. Furthermore, in the point-to-point case, parallel links can at most sum up their individual capacities, therefore
\begin{align}
C(T, \bN^{(1)}, \bN^{(2)} ) &\leq \min(R^{(1)}, R^{(2)})\\
R^{(1)} = \sum_{i=1}^{L_{s,r}} R^{(1)}_i &= \sum_{i=1}^{L_{s,r}} C(T - N^{(2)}_{\textrm{min}}, N^{(1)}_i)=\frac{L_{s,r}(T-N^{(2)}_{\textrm{min}}+1)-\sum_{i=1}^{L_{s,r}}N^{(1)}_i}{T-N^{(2)}_{\textrm{min}}+1} \label{eq:ratefirsthop}\\
R^{(2)} = \sum_{i=1}^{L_{r,d}} R^{(2)}_i &= \sum_{i=1}^{L_{r,d}} C(T - N^{(1)}_{\textrm{min}}, N^{(2)}_i)=\frac{L_{r,d}(T-N^{(1)}_{\textrm{min}}+1)-\sum_{i=1}^{L_{r,d}}N^{(2)}_i}{T-N^{(1)}_{\textrm{min}}+1} \label{eq:ratesecondhop}
\end{align}

Denoting 
\begin{align}
    \bar{N}^{(1)} = \frac{1}{L_{s,r}} \sum_{i=1}^{L_{s,r}} N^{(1)}_i,~
    \bar{N}^{(2)} = \frac{1}{L_{r,d}} \sum_{i=1}^{L_{r,d}} N^{(2)}_i,
    \label{eq:N_avg}
\end{align}
we have
\begin{align}
    C(T, \bN^{(1)}, \bN^{(2)} ) &\leq \min\left(\frac{T-N^{(2)}_{\textrm{min}}+1-\bar{N}^{(1)}}{\frac{1}{L_{s,r}}(T-N^{(2)}_{\textrm{min}}+1)},\frac{T-N^{(1)}_{\textrm{min}}+1-\bar{N}^{(2)}}{\frac{1}{L_{r,d}}(T-N^{(1)}_{\textrm{min}}+1)}\right)
    \label{eq:explicit_upper}
\end{align}

\begin{remark}
    Note that, if $T - N^{(2)}_{\textrm{min}} - N^{(1)}_i \leq 0$, the $i$th link of the first hop has rate 0 and therefore it can be discarded. The same holds for $T - N^{(1)}_{\textrm{min}} - N_i^{(2)} \leq 0$, in which the $i$th link of the second hop may be discarded. Therefore, we may focus our analysis on the scenario where 
    \begin{equation}
        T \geq T_{\textrm{min}} = \max\left( \max(\bN^{(1)}) + \min(\bN^{(2)}), \max(\bN^{(2)}) + \min(\bN^{(1)})\right). \label{eq:tmin}
    \end{equation}
\end{remark}
\begin{example}
	In the network described in Fig.\ref{fig:achievex}, we have
	\begin{align}
	R^{(1)}_1 = C(4, 2) = \frac{3}{5}, \quad R^{(1)}_2 = C(4, 3) = \frac{2}{5}, \quad
	R^{(2)}_1 = C(3, 1) = \frac{3}{4}, \quad
	R^{(2)}_2 = C(3, 2) = \frac{2}{4} 
	\end{align}
	thus $R^{(1)} = 1$ and $R^{(2)} = 5/4$, finally, $C \leq 1$. But since we have shown a coding scheme that achieves $R = 1$, we have $C = 1$. 
	
	This is an interesting example, as it shows that even an upper bound that might seem very optimistic can be tight, and a coding scheme that is sufficiently simple can be capacity-achieving. Unfortunately, as will be seen, this is not true in general, but still holds for a wide range of parameters, as shown in Section~\ref{sec:numerical}.
\end{example}









\section{Baseline Schemes}\label{sec:baseline}

Before we propose our scheme, we would like to present the schemes against which we will be comparing our scheme, i.e., baseline schemes that provide possibly the simplest solutions to the multi-link setting problem. 

\subsection{Message-wise decode-and-forward}

In this baseline scheme, all symbols are recovered at the same time $T^{(1)}_{mw}$ by the relay, decoded and re-encoded, and then recovered at time $T$ by the destination. As in \cite{Silas2019}, we denote this strategy a ``message-wise'' decode-and-forward. Under this scenario, random linear codes achieve optimal rate, as random linear codes are point-to-point optimal \cite{badr2017layered}. It is also fairly simple to derive a maximum achievable rate for this scheme, as follows: let us denote $T^{(2)}_{mw} = T - T^{(1)}_{mw}$. Then, the rate in each hop is bounded by
\begin{align}
    (T^{(1)}_{mw} + 1)k \leq \sum_{i=1}^{L_{sr}} (T^{(1)}_{mw} + 1 - N^{(1)}_i) n^{(1)}_i, \quad \textrm{and} \quad (T^{(2)}_{mw} + 1)k \leq \sum_{i=1}^{L_{sr}} (T^{(2)}_{mw} + 1 - N^{(2)}_i) n^{(2)}_i.
\end{align}
This can be seen from a simple counting argument: the number of channel symbols should be larger than or equal to the number of information symbols we wish to recover. 

Due to the definition of rate we used, we are able to obtain two constraints on the achievable rate of this scheme
\begin{align}
    \frac{k}{n} \leq \frac{ \sum_{i=1}^{L_{sr}} (T^{(1)}_{mw} + 1 - N^{(1)}_i)}{T^{(1)}_{mw} + 1}, \quad \textrm{and} \quad \frac{k}{n} \leq \frac{ \sum_{i=1}^{L_{sr}} (T^{(2)}_{mw} + 1 - N^{(2)}_i)}{T^{(2)}_{mw} + 1}.
\end{align}

Finally, we may optimize $T^{(1)}_{mw}$ and $T^{2)}_{mw}$ to find the highest achievable rate this scheme can provide, thus, the rate is given by
\begin{align}
\frac{k}{n} = \max_{T^{(1)}_{mw}+T^{(2)}_{mw}\leq T} \quad & \min \left(\frac{ \sum_{i=1}^{L_{sr}} (T^{(1)}_{mw} + 1 - N^{(1)}_i)^+}{T^{(1)}_{mw} + 1},  \frac{ \sum_{i=1}^{L_{sr}} (T^{(2)}_{mw} + 1 - N^{(2)}_i)^+}{T^{(2)}_{mw} + 1}       \right)
\end{align}

\begin{remark}
    In this scheme, it suffices to perform separate coding in each link, as it is easy to see that joint encoding across all links does not improve the achievable rate.
\end{remark}



\subsection{Concatenated symbol-wise decode and forward} \label{sec:concSW}

First, let us consider the scenario where $L_{s, r} = 1$. Each link in the relay-destination hop transmits at its upper bound rate. In order to do that, we set $k^{(2)}_i = T - N^{(1)} - N^{(2)}_i + 1$ and $n^{(2)} = n^{(2)}_i = T - N^{(1)} + 1$. Note that $R^{(2)}_i = k^{(2)}_i/n^{(2)}_i$ is the same as \eqref{eq:explicit_upper}. Then, the link in the first hop uses a concatenation of the codes for each independent link in the second hop, that is, $k^{(1)} = k = \sum_{i=1}^{L_{r,d}} T - N^{(1)} - N^{(2)}_i + 1$ and $n^{(1)} = \sum_{i=1}^{L_{r,d}} T + 1 - N_i^{(2)}$. Thus, the achievable rate is given by $R_{\rm con-SW} \geq \frac{k}{\max(n^{(1)}, n^{(2)})}$.

In order to generalize such scheme, we consider all possible sub-networks and concatenate them together. For example, consider the network in Fig.~\ref{fig:achievex}. We first consider the sub-network with $N^{(1)} = 2$. In this case, the rates are $R^{(2)}_{1,1} = \frac{3}{4}$, $R^{(2)}_{2,1} = \frac{2}{4}$ and $R^{(1)}_1 = \frac{3 + 2}{5 + 4} = \frac{5}{9}.$ Similarly, for $N^{(1)} = 3$, we would get $R^{(2)}_{1,2} = \frac{2}{3}$, $R^{(2)}_{2,2} = \frac{1}{3}$ and $R^{(1)}_2 = \frac{2 + 1}{5 + 4} = \frac{3}{9}$. Finally, we concatenate both codes for the second hop, getting $R^{(2)}_1 = \frac{3 + 2}{4 + 3} = \frac{5}{7}$ and $R^{(2)}_2 = \frac{2 + 1}{4 + 3} = \frac{3}{7}$. Finally, we have $R^{(1)} = \frac{3 + 5}{9}$ and $R^{(2)} = \frac{5 + 3}{7}$. The minimum is, evidently, $R^{(1)}$ and this scheme achieves rate $8/9$.

\begin{proposition}\label{prop:concSWrate}
    This concatenated symbol-wise decode and forward scheme achieves a rate lower bounded by
    \begin{equation}
      R_{\rm con-SW}\geq  \frac{T+1-\bar{N}^{(2)}-\bar{N}^{(1)}}{\max\left(\frac{1}{L_{s,r}}\left(T-\bar{N}^{(2)}+1\right),\frac{1}{L_{r,d}}\left(T-\bar{N}^{(1)}+1\right)\right)}
    \label{eq:explicit_R_con_SW}
    \end{equation}
\end{proposition}

\begin{remark}
    The rate in \eqref{eq:explicit_R_con_SW} is always achievable, however, as can be seen in the proof, it can be improved by discarding links which result in a zero rate. However, in that case, we do not obtain a closed-form expression.
\end{remark}

\begin{proposition}\label{prop:concSWoptimal}
    This concatenated symbol-wise scheme achieves the upper bound if and only if the maximal number of erasures in all links of the hop that is not the bottleneck are equal.
\end{proposition}


\section{Scheme Proposal}\label{sec:scheme}

We begin explaining our scheme for a particular scenario where $L_{s, r} = 1$. Later, we generalize the scheme for a general $L_{s,r}$. We split our scheme proposal in two subsections. In the first, we present a greedy algorithm that aims to achieve the proposed upper bound. However, in general, our  decode-and-forward separate coding strategy is unable to achieve the upper bound to the best of our knowledge, and, in these cases, we wish to improve the rate in order to get closer to the upper bound. This improvement is presented in the second subsection. Then, a brief subsection generalizes the scheme for a general $L_{s,r}$.

\subsection{Initial iteration - aiming for the upper bound}\label{sec:initialit}

We start by computing $R^{(1)}$ and $R^{(2)}$ as in \eqref{eq:ratefirsthop} and \eqref{eq:ratesecondhop}. We then find which one is the bottleneck, i.e., which one is the lower rate. If both rates are equal, we choose the bottleneck to be the hop in which the sum of the number of erasures is larger. For this analysis, we will assume the first hop is the bottleneck. The other case will be included in the general case.

For all links, we set $$n^{(1)}_i = n^{(2)}_j = n = (T + 1 - N^{(2)}_{\textrm{min}})(T + 1 - N^{(1)}).$$ Then, we set the number of symbols to be transmitted in the first hop as
\begin{align}
k^{(1)} = R^{(1)} \cdot n.
\end{align}
Note that, although at first glance this is the upper bound rate in \eqref{eq:ratefirsthop}, this may not be achievable. In that case, some of the symbols carry no information, e.g., are set to trivial symbols (0). 

We then compute the worst delay for that hop as
\begin{align}
T^{(1,g)}[1] = \left\lceil \frac{N^{(1)} n}{n - k^{(1)}} - 1 \right\rceil \overset{(a)}{=} T - N_{\textrm{min}}^{(2)}
\end{align}
where $(a)$ follows from definition of $k^{(1)}$. The list of possible delays is then
\begin{align}
\bT^{(1, g)} =  [T - N_{\textrm{min}}^{(2)}, T - N_{\textrm{min}}^{(2)}-1, \ldots, N^{(1)} + 1, N^{(1)}].
\end{align}
That way, the tuples for the first hop are given by
\begin{align}
        {\small
        \begingroup 
        \setlength\arraycolsep{1.5pt}
            \bG^{(1)} = \left[\begin{matrix} 
            \left(T - N_{\textrm{min}}^{(2)}, n - \frac{T - N_{\textrm{min}}^{(2)}}{N^{(1)}}(n - k^{(1)})    \right),&\left(T - N_{\textrm{min}}^{(2)} - 1, \frac{n - k^{(1)}}{N^{(1)}}    \right),& \cdots ,& \left(N^{(1)} + 1, \frac{n - k^{(1)}}{N^{(1)}}  \right),& \left(N^{(1)}, \frac{n - k^{(1)}}{N^{(1)}}  \right)
            \end{matrix}\right]
        \endgroup
        }
        \end{align}
where the number of symbols in each delay is the maximum achievable as seen in Section~II. Finally, we can compute the overall delay spectrum of the $L$-links code employing separate coding strategy using these single-link codes. The number of symbols in each delay is then given by
\begin{align}
k^{(1)}[1] = n - \frac{T - N_{\textrm{min}}^{(2)}}{N^{(1)}}(n - k^{(1)})~;~
k^{(1)}[j] = \frac{n - k^{(1)}}{N^{(1)}} \forall j \geq 2
\end{align}

Then, we apply the following idea for the remaining hop: if $k^{(1)}[j]$ symbols can be transmitted with a delay $T^{(1)}[j]$ through the bottleneck, then at most $k^{(1)}[j]$ symbols can be transmitted with a delay $T - T^{(1)}[j]$ in the second hop. However, even constraining our coding scheme to separate coding, it is still unclear how to distribute the number of symbols in each link. From empirical analysis, we have noticed that maximizing the rate of the links with the worst number of erasures is beneficial, as it allows for a better delay spectrum in the links with less erasures. For that reason, we propose the following heuristic approach to solving this optimization:
\begin{itemize}
	\item Initialization:
	\begin{enumerate}
		\item Order the links in the second hop (or, generally, in the hop that is not the bottleneck) in a decreasing order of number of erasures, that is, in such a way that $N^{(2)}_1 \geq N^{(2)}_2 \geq \cdots \geq N^{(2)}_{L_{r,d}}$.
		\item Generate the following tuple constraint:
		\begin{align*}
        {
        \begingroup 
        \setlength\arraycolsep{1.5pt}
            \bG^{\textrm{con}} = \left[\begin{matrix} \left( T - N^{(1)} , k^{(1)}[\ell^{(1, g)}] \right),&\left( T - N^{(1)} - 1 , k^{(1)}[\ell^{(1, g)}-1] \right),&\cdots,&\left( N_{\textrm{min}}^{(2)} , k^{(1)}[1] \right),&\left( N_{\textrm{min}}^{(2)} - 1 , 0 \right) \end{matrix}\right]
        \endgroup
        }
        \end{align*}

		where $\ell^{(1, g)} = T + 1 - N^{(1)} - N^{(2)}_{\textrm{min}}$. Note that, for convenience, it is ordered starting by the largest delay and contains an extra element which states that no symbols can be transmitted at a smaller delay than $N^{(2)}_{\textrm{min}}$. Then, let $\bk^{\textrm{con}} = [k^{(1)}[\ell^{(1, g)}] , k^{(1)}[\ell^{(1, g)}-1] , \ldots, k^{(1)}[2], k^{(1)}[1], 0]$. 
	\end{enumerate}
	\item For each link $i$ from $1$ to $L_{r,d}$ in the second hop, do:
	\begin{enumerate}
		\item Define a vector $\bT_i$ consisting of the elements in the first position (i.e., respective to the tuple) of $\bG^{\textrm{con}}$, i.e.,
		\begin{equation}
		\bT_i = [T - N^{(1)}, \ldots, N_i^{(2)}, N_i^{(2)} - 1, \ldots, N^{(2)}_{\textrm{min}} - 1] \label{eq:DelayAlg}
		\end{equation}
		\item Define a vector $\bR^{\textrm{con}}$, where each element is defined as $\bR^{\textrm{con}}[j] = \sum_{\ell = 1}^{j - 1} \frac{k^{\textrm{con}}[j] }{n}$
		\item Solve, for $j$ from $1$ to $T - N^{(1)} - N_i^{(2)} + 2$
		\begin{align}
		\bk'[j] = n - n \cdot N^{(2)}_i \left( \frac{1 - \bR^{\textrm{con}}[j]}{\bT_i[j] + 1} \right). \label{eq:DelayConLow}
		\end{align}
		\item Set $k^{(2)}_i = \left\lfloor   \min(\bk') \right\rfloor$. This is the maximum number of symbols that may be transmitted constrained by the number of symbols allowed in each delay, as seen in Corollary~\ref{corollary:maxsym}, that is, except for the floor operation, this is optimal.
		\item If $\frac{n - k^{(2)}_i}{N^{(2)}_i}$ is not an integer, multiply $n$ and all $k^{(2)}_{i'}$, $i'$ from $1$ to $i$, by $N^{(2)}_i$. Multiply $\bk^{\textrm{con}}$ by $N^{(2)}_i$.
		\item Compute the number of symbols transmitted in each delay as in \eqref{eq:bestsymbols}. Denote the vector as $\bk^{(2)(g)}$, pairing with the appropriate delays.
		\item Update $k^{\textrm{con}}$ based on how many symbols were transmitted with each delay through this link, that is, subtract $\bk^{(2)(g)}$ from $k^{\textrm{con}}$, correctly pairing the respective delays. If there is any negative entry, that value is moved to the next larger delay available.
		\item Move on to $i+1$. 
	\end{enumerate}
	\item Finally, match the delay spectrum of the second hop with the delay spectrum of the first hop. If the number of symbols in the bottleneck is larger than in the other hop, we remove the symbols with largest delay (e.g. transmit zeros in those positions) in the bottleneck.
\end{itemize}



A detailed example of this first iteration is given for a three node network with total delay of 4 consisting of a single link between the source and relay (subjected to maximal of a single erasure) and two links between the relay and destination (one subjected to a maximal of three erasures and the other subjected to a maximum of two erasure) which is depicted    Fig.~\ref{fig:1t2ex} and detailed in Appendix~\ref{ex:firststep} (denoted as Example~3).


\begin{figure}[h!]
\centering
\includegraphics[]{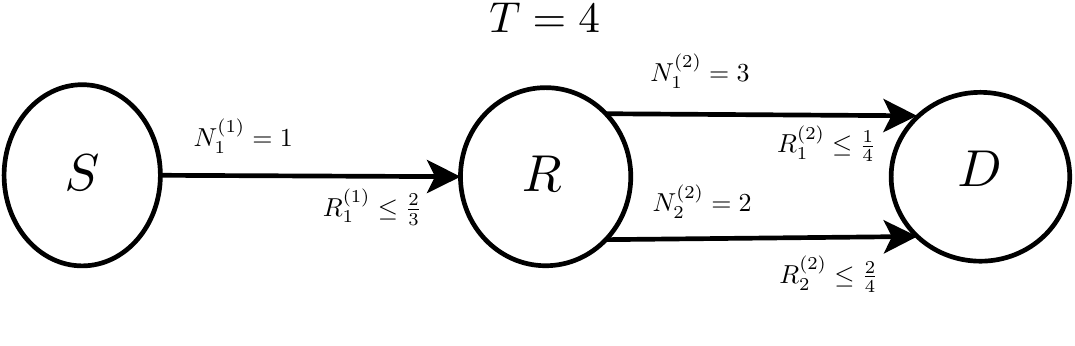}
\caption{One-to-two relay setting.}
\label{fig:1t2ex}
\end{figure}

\subsection{Additional iterations - gradually lowering the target} \label{sec:optimizationbn}

The previous algorithm is not able to achieve the upper bound in some scenarios, as shown in Example~\ref{ex:firststep}. In such scenarios, there is a gap between the number of symbols transmitted in the bottleneck and the number of symbols transmitted in the other hop. In fact, we arrive at a contradiction: the hop we assumed to be the bottleneck operates at a ``higher rate'' than the other hop. Previously, we simply erased such symbols, making them trivial symbols. However, such choice is certainly suboptimal: instead, by transmitting less symbols, we may be able to achieve a better delay spectrum for the remaining symbols. As a simple example, consider a rate $9/12$ code under $1$ erasure. This code transmits $3$ symbols with each delay $[1, 2, 3]$, i.e., a concatenation of three $3/4$ codes. If we wish to reduce it to a $8/12$ code, by erasing 1 symbol with delay 3, we achieve 3 symbols transmitted with delay 1, 3 symbols with delay 2, and 2 symbols with delay 3. However, we can, instead, change the code to four concatenations of a $2/3$ code, achieving 4 symbols with delay 1 and 4 symbols with delay 2, clearly a better choice. For this reason, we wish to increase the achieved rate in a smarter way.

Maintaining the assumption that the first hop is the bottleneck, having $k^{(1)} > k^{(2)}$ means we should decrease $k^{(1)}$. The general principle is:
\begin{itemize}
	\item If $k^{(1)} > k^{(2)}$, decrease $k^{(1)}$, while maintaining it higher than the previously found $k^{(2)}$.
	\item If $k^{(1)} = k^{(2)}$, increase $k^{(1)}$. 
\end{itemize}
In the first scenario, the rate is guaranteed to, at least, stay the same, as decreasing $k^{(1)}$ can only make the delay spectrum in the first hop better, thus, the same previously found delay spectrum for the second hop is still achievable. The second scenario may decrease the rate. Therefore, we are interested in finding the maximum $k^{(1)}$ such that $k^{(1)} = k^{(2)}$. In order to solve this problem, we use a simple heuristic together with a bisection algorithm. For the first point of operation, we choose $k^{(1)}$ and $n$ satisfying the rates found in Section~\ref{sec:concSW}. Afterwards, we continue the optimization using a bisection algorithm, that is, if $k^{(1)} > k^{(2)}$, we update the upper bound to $k^{(1)}$, otherwise, we update the lower bound to $k^{(1)}$. The subsequent point of operation is given by the average between the upper bound and lower bound. If required, we multiply all parameters by 2, in order to allow smaller steps in the algorithm. We stop the algorithm once the upper bound and lower bound are close enough. 

It should be noted that this algorithm may result in extremely large codes, with $n$ larger than, for example, $10^5$. If such codes are impractical, additional breaking conditions may be included, such as a maximum $n$ or a maximum number of iterations, at the cost of a small loss in performance. 

As a motivation for our algorithm, we present an example comparing the achievable rate using the greedy algorithm and then update the bottleneck using our adaptation, considerably improving the rate. 

Consider the setting in Fig.~\ref{fig:1t2ex}. Running our greedy algorithm, we obtain $n = 12$, $k^{(2)} = 7$ and $k^{(1)} = 8$. We then set the upper bound to be $8/12$ and the lower bound to be $7/12$. Since we do not achieve the upper bound, we use the parameters from Section~\ref{sec:concSW}, that is, $k^{(1)} = 3$ and $n = 5$. This results in rate $0.6$, slightly higher than $7/12$. We then update our lower bound to this value. Recall that the upper bound is $2/3$, so, after achieving 0.6, we attempt 0.6333. This is achieved with equality (i.e., $k^{(1)} = k^{(2)}$, thus we again update it to 0.65, which is achieved with equality again. We then update $k^{(1)}$ to 0.6583, however, we now notice a gap (i.e., $k^{(1)} > k {(2)}$), and the achieved $k^{(2)}$ is lower than 0.65. Thus, we set the new upper bound to this value. After a few more iterations, the algorithm converges to 0.65 (i.e., all points above it are achieved with some gap, and the achieved rate is always at most 0.65).

\subsection{General $L_{s,r}$}
We now assume, without loss of generality, that the first hop is the bottleneck. This is now without loss of generality because our scheme is symmetrical (as is the upper bound) with respect to the hops. First, let us generalize Section~\ref{sec:initialit}.

In this case, we have 
\begin{equation}
    n^{(1)}_i = n^{(2)}_j = n = (T + 1 - N^{(2)}_{\textrm{min}})(T + 1 - N^{(1)}_{\textrm{min}}) \quad \textrm{and} \quad k^{(1)}_i = R^{(1)}_i \cdot n.
\end{equation}

Furthermore, the worst delay is now computed for each link, although it is the same for all of them, as
\begin{align}
T^{(1,g)}_i[1] = \left\lceil \frac{N^{(1)}_i n}{n - k^{(1)}_i} - 1 \right\rceil \overset{(a)}{=} T - N_{\textrm{min}}^{(2)} 
\end{align}
where $(a)$ follows from definition of $k_i^{(1)}$. Similarly, the list of possible delays for each link is different, and computed as
\begin{align}
\bT^{(1, g)}_i =  [T - N_{\textrm{min}}^{(2)}, T - N_{\textrm{min}}^{(2)}-1, \ldots, N^{(1)}_i + 1, N^{(1)}_i]. \label{eq:delaygroups}
\end{align}
Note that, since every link actually has the same $T_i^{(1, g)}$, which is $T - N_{\textrm{min}}^{(2)}$, we can further extend the possible delays for all links and make all the links have a common ``possible'' delay spectrum, making computation easier later. That way, the tuples for each link are given by

\begin{align}
{\footnotesize
\begingroup 
\setlength\arraycolsep{1.5pt}
    \bG^{(1)}_i = \left[\begin{matrix} \left(T - N_{\textrm{min}}^{(2)}, n - \frac{T - N_{\textrm{min}}^{(2)}}{N^{(1)}_i}(n - k^{(1)}_i) \right), & \left(T - N_{\textrm{min}}^{(2)} - 1, \frac{n - k^{(1)}_i}{N^{(1)}_i} \right) 
    ,& \cdots ,& \left(N^{(1)}_i, \frac{n - k^{(1)}_i}{N^{(1)}_i}  \right)  ,& \left(N^{(1)}_i - 1, 0 \right), & \cdots ,& \left(N^{(1)}_{\textrm{min}}, 0   \right) \end{matrix}\right]
\endgroup
}
\end{align}

Finally, the computation of the overall delay spectrum of the $L$-links code differs from the case where $L_{s,r} = 1$. Employing separate coding strategy using these single-link codes, the number of symbols in each delay is given by
\begin{align}
k^{(1)}[1] = \sum_{i=1}^{L_{s,r}} n - \frac{T - N_{\textrm{min}}^{(2)}}{N^{(1)}_i}(n - k^{(1)}_i)~;~
k^{(1)}[j] = \sum_{i : N^{(1)}_i \leq T - N_{\textrm{min}}^{(2)} - j + 1   }  \frac{n - k^{(1)}_i}{N^{(1)}_i} \forall j \geq 2
\end{align}
which, although a more complicated expression, can still be easily computed.

With that, the constraint tuples change to
\begin{align}
{
\begingroup 
\setlength\arraycolsep{1.5pt}
\bG^{\textrm{con}} = \left[
\begin{matrix} \left( T - N^{(1)}_{\textrm{min}} , k^{(1)}[\ell^{(1, g)}] \right),&\left( T - N^{(1)}_{\textrm{min}} - 1 , k^{(1)}[\ell^{(1, g)}-1] \right),&,\cdots,&\left( N_{\textrm{min}}^{(2)} , k^{(1)}[1] \right) ,&\left( N_{\textrm{min}}^{(2)} - 1 , 0 \right)
\end{matrix}\right]
\endgroup
}
\end{align}
the delays computed for each link of the algorithm in \eqref{eq:DelayAlg} change to
\begin{equation}
\bT_i = [T - N^{(1)}_{\textrm{min}}, \ldots, N_i^{(2)}, N_i^{(2)} - 1, \ldots, N^{(2)}_{\textrm{min}} - 1]
\end{equation}
and \eqref{eq:DelayConLow} must be solved for $j$ from $1$ to $T - N^{(1)}_{\textrm{min}} - N_i^{(2)} + 2$.


A full example on the generalized algorithm achieving the upper bound in the setting from Fig.~\ref{fig:achievex} can be found in the Appendix~\ref{ex:algorithm} as Example~2.

As for Section~\ref{sec:optimizationbn}, a major change occurs: now, it is unclear how we should update $k^{(1)}$, i.e., how many symbols we should remove from or add to each link. Based on the same intuition that we should allow the link with less erasures have a lower rate in order to achieve better delay spectra, we propose the following heuristic:
\begin{enumerate}
	\item Order the links in decreasing number of erasures.
	\item Decrease or increase $R_1^{(1)}$, with a maximum rate of $R_1^{(1)} \leq \frac{T + 1 - N^{(1)}_1 - N^{(2)}_{\textrm{min}}}{T + 1 - N^{(2)}_{\textrm{min}}} $.
	\item If $R_i^{(1)}$ achieves its upper bound, and we need to further increase the rate, we fix it at its upper bound and start changing $R_{i+1}^{(1)}$.
\end{enumerate}

Note that, since the first point of operation is known to be achievable from Section~\ref{sec:concSW}, the rate should only increase from that point.

Finally, let us formally put the lemmas together to show that this scheme works.

\begin{theorem}\label{theorem:achievability}
	Under $\bN^{(1)}$ erasures in the first hop and $\bN^{(2)}$ erasures in the second hop, there exists a code with our choice of parameters that allows the destination to recover all source symbols by the deadline.
\end{theorem}

Unfortunately, for the general case, a closed form expression is hard to derive. However, we conjecture that the rate our optimization scheme achieves is always at least that of the baseline schemes. Such conjecture is based on the fact that we use upper-bound-achieving codes in each link and that the rate achieved by the random linear codes does not increase by not employing a separate coding strategy, and the fact that, when we do not achieve the upper bound, we use the rate of the concatenated symbol-wise scheme as our starting point for optimization. Thus, given the constraint of symbol-wise decode-and-forward and separate coding, the only source of non-optimality in our scheme is the way we compute the rates used in each link. However, as we shall see in the following subsection, empirical evidence strongly suggests that our conjecture holds.

\subsection{Numerical results}\label{sec:numerical}

\begin{figure}
    \centering
    \begin{subfigure}{.45\textwidth}
        \centering
        \includegraphics[trim=3.5cm 8.5cm 4cm 8.5cm, clip, width=\textwidth]{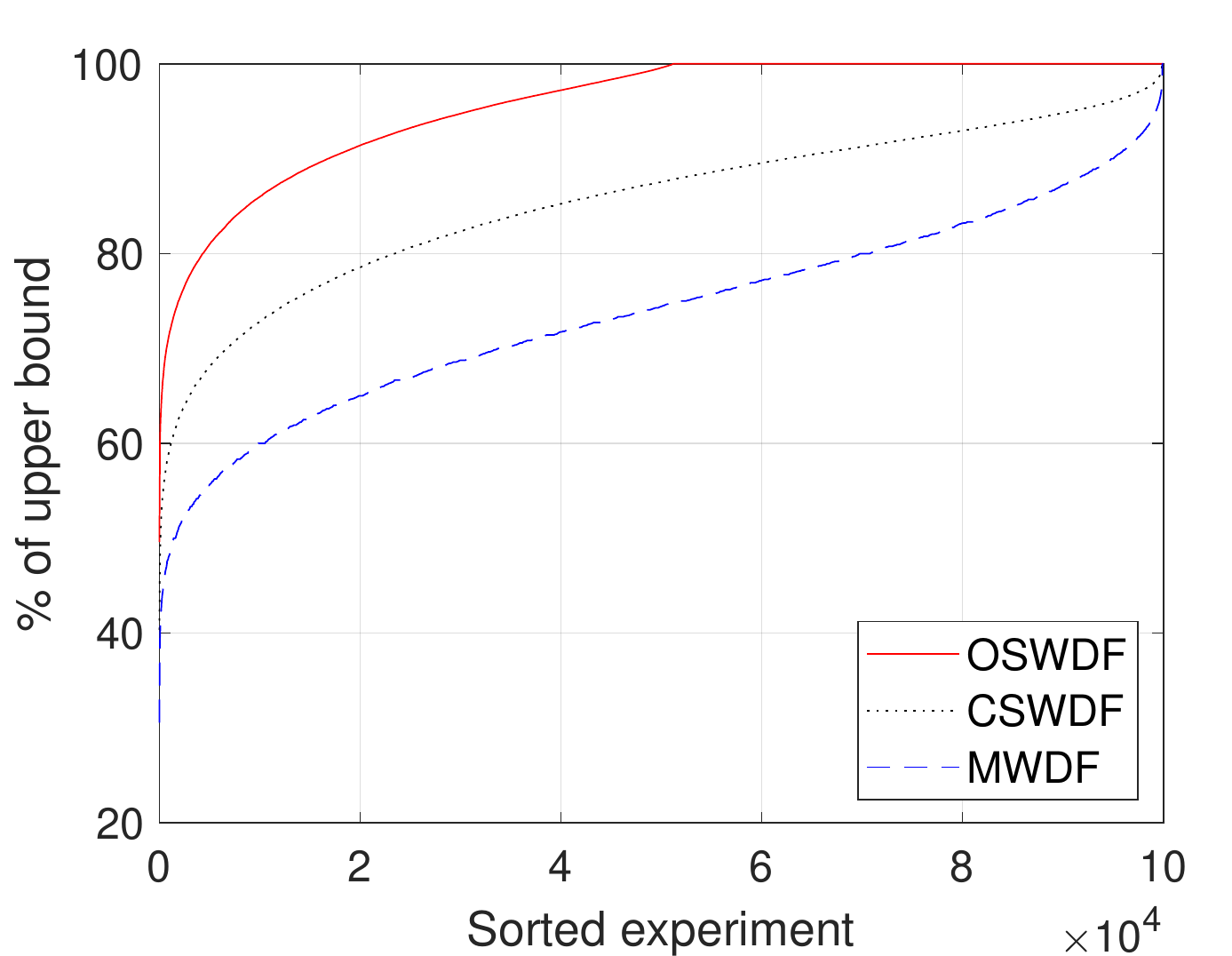}
        \caption{Comparison against upper bound.}\label{fig:upperboundcomparison}
    \end{subfigure}\hspace{.2cm}
    \begin{subfigure}{.45\textwidth}
        \centering
        \includegraphics[trim=3.5cm 8.5cm 4cm 8.5cm, clip, width=\textwidth]{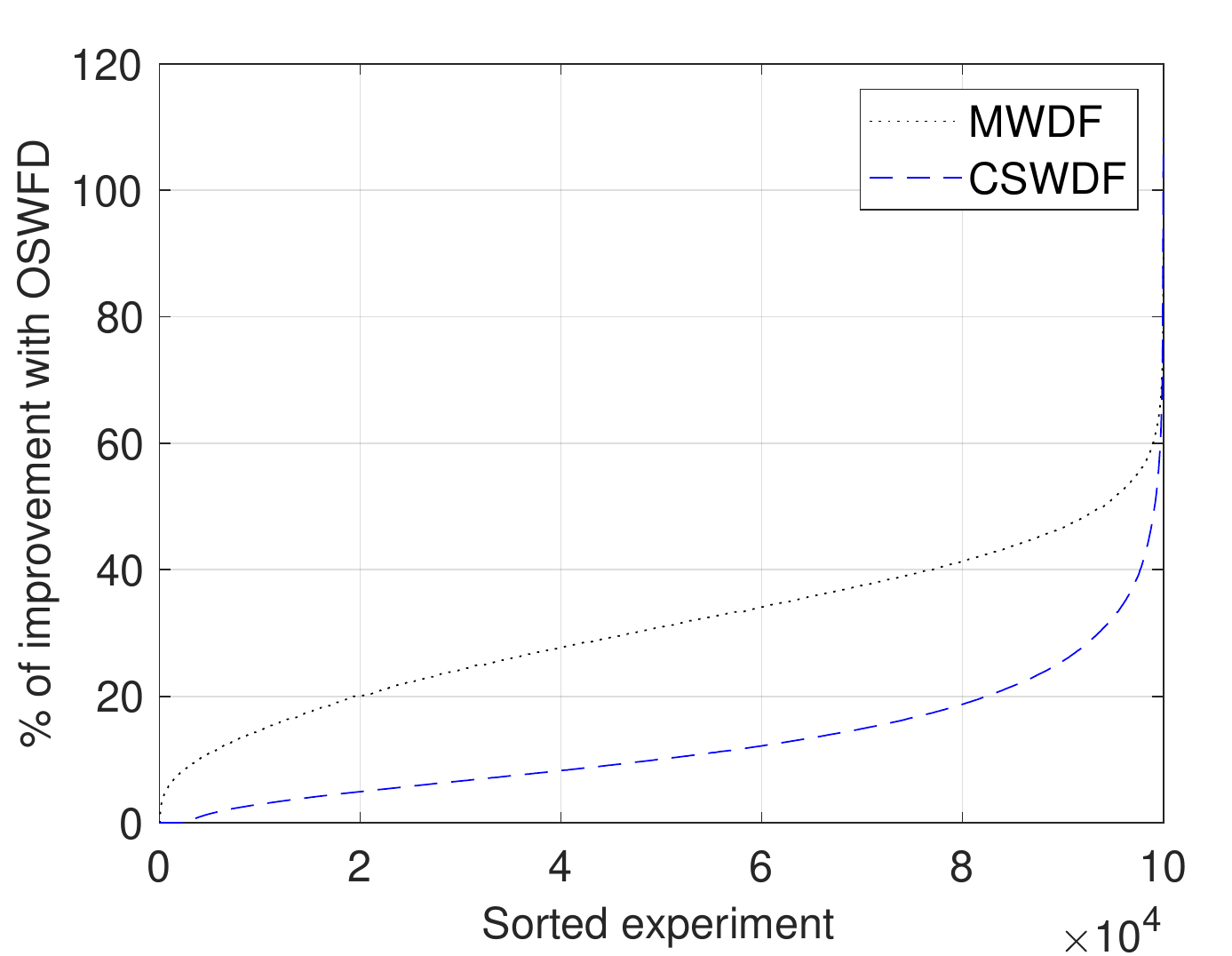}
        \caption{Improvement using OSWDF.}\label{fig:directcomparison}
    \end{subfigure}\hspace{.2cm}
    \caption{Comparison between achieved rate by proposed scheme against baseline schemes.}
    \label{fig:cdf}
\end{figure}

In order to compare the rates achieved by the proposed scheme and the baseline schemes, we have computed several scenarios with different number of links, erasures and delays. The following experiment has been done:
\begin{itemize}
    \item We set the number of links in both the source-relay hop and in the relay-destination hop as a random integer between 3 and 6.
    \item We set the number of erasures in each link as a random integer between 1 and 10.
    \item We set the delay as a random integer between $T_{\textrm{min}}$ and $T_{\textrm{min}} + 10$, where $T_{\textrm{min}}$ is as in \eqref{eq:tmin}.
\end{itemize}
Then, we run such experiment $10^5$ times. Fig.~\ref{fig:cdf} presents sorted curves comparing our scheme against the baseline schemes. In Fig.~\ref{fig:directcomparison}, it can be seen that, in no experiments any of the baseline schemes has achieved a higher rate than our scheme. Compared to the MWDF scheme, our scheme achieves significantly higher rates ($\geq 30\%$ improvement) half the time. Compared to the concatenated symbol-wise scheme, the improvement in the rate is more modest most of the time, but there is still a considerable number of scenarios where our scheme is able to outperform it considerably. However, a more interesting behavior can be seen in Fig.~\ref{fig:upperboundcomparison}. While our scheme achieves the upper bound in about half of the experiments, the baseline schemes rarely are able to.


\section{Propagation Delay}\label{sec:propdelay}

In order to further approximate our model to actual channels, we wish to consider differing propagation delays in each link. As in the erasure pattern, we model it considering a worst case scenario, thus allowing us to fix the propagation delays and obtaining useful expressions. In order to model the propagation delay, we change the definition of our channel. Instead of an erasure channel, as defined in Definition~\ref{def:channel}, we now consider a delayed erasure channel. All definitions from Definition~\ref{def:channel} are kept the same, except for \eqref{eq:erasurerelay} and \eqref{eq:erasuredest}. These are respectively changed to
$\by^{(1)}_{t, i} = g_{n_i^{(1)}}(\bx_{t - \Delta T_i^{(1)}, i}^{(1)}, e_{t, i}^{(1)})$ and $\by^{(2)}_{t, i} = g_{n_i^{(2)}}(\bx_{t - \Delta T_i^{(2)}, i}^{(2)}, e_{t, i}^{(2)})$, where $\Delta T^{(h)}_i$ is the propagation delay in the $i$th link in the $h$th hop.

In this case, the intuition is fairly simple: the effective delay allowed in each link is reduced by its respective propagation delay. Below, we summarize the changes that need to be done in the previous expressions and algorithm.

Let us denote by $\mathbf{\Delta T}^{(1)}$ the propagation delay vector associated with the first hop and $\mathbf{\Delta T}^{(2)}$ the propagation delay vector associated with the second hop. Let us denote by $\bZ^{(1)} = \bN^{(1)} +  \mathbf{\Delta T}^{(1)} $ and $\bZ^{(2)} = \bN^{(2)} +  \mathbf{\Delta T}^{(2)}$. Finally, let us denote by $Z_{\textrm{min}}^{(1)} = \min(\bZ^{(1)})$ and $Z_{\textrm{min}}^{(2)} = \min(\bZ^{(2)})$.

Equations \eqref{eq:ratefirsthop} and \eqref{eq:ratesecondhop} change, respectively, to $R^{(1)} \leq \sum_{i = 1}^{L_{s,r}} R_i^{(1)} = \sum_{i=1}^{L_{s,r}} C(T - Z_{\textrm{min}}^{(2)} - \Delta T_i^{(1)}, N_i^{(1)})$ and $R^{(2)} \leq \sum_{i = 1}^{L_{r,d}} R_i^{(2)} = \sum_{i=1}^{L_{r,d}} C(T - Z_{\textrm{min}}^{(1)} - \Delta T_i^{(2)}, N_i^{(2)})$.
We note that these rates are strictly larger than modeling the propagation delay as more erasures, that is, $C(T - Z_{\textrm{min}}^{(2)} - \Delta T_i^{(1)}, N_i^{(1)}) > C(T - Z_{\textrm{min}}^{(2)}, N_i^{(1)} + \Delta T_i^{(1)})$, therefore such simplification would lead to certainly suboptimal codes. Further, due to the new upper bound expressions, the condition for the minimum $T$ such that all links can be active (i.e., are not certainly discarded) is given by $$T \geq \max \left( \max(\bZ^{(1)}) + \min(\bZ^{(2)}) , \max(\bZ^{(2)}) + \min(\bZ^{(1)})\right).$$

In the baseline scheme, similarly, the number of recovered symbols in each link changes to
\begin{align}
    R^{(1)}_{mw} = \sum_{i=1}^{L_{s,r}} \frac{ (T^{(1)}_{mw} - \Delta T_i^{(1)} + 1 - N^{(1)}_i)^+}{T^{(1)}_{mw} - \Delta T_i^{(1)} + 1},~
    R^{(2)}_{mw} = \sum_{i=1}^{L_{r,d}} \frac{  (T^{(2)}_{mw} - \Delta T_i^{(2)} + 1 - N^{(2)}_i)^+}{T^{(2)}_{mw} - \Delta T_i^{(2)} + 1}
\end{align}

As for the optimization algorithm, the following changes must be done:
\begin{itemize}
    \item The ordering of the links should be done according to $\bZ^{(1)}$ and $\bZ^{(2)}$, rather than $\bN^{(1)}$ and $\bN^{(2)}$.
    \item When computing the tuples for delay and number of symbols, the tuple should take into account the delay propagation of each link. For example, \eqref{eq:delaygroups} changes to
    \begin{equation}
        \bT_i^{(1, g)} = \left[ T - Z_{\textrm{min}}^{(2)}, T - Z_{\textrm{min}}^{(2)}, \ldots, Z_i^{(1)} + 1, Z_i^{(1)}\right].
    \end{equation}
    Note, however, that the denominator term in the number of symbols, e.g., $N_i^{(1)}$ in $\frac{n - k^{(1)}_i}{N_i^{(1)}}$, does not change, reinforcing the idea that modeling the propagation delay as extra erasures is a pessimistic model.
    \item The starting $n$ changes to $n = (T + 1 - Z_{\textrm{min}}^{(2)})(T + 1 - Z_{\textrm{min}}^{(1)})$.
\end{itemize}

\section{Simulations}\label{sec:simulation}

As mentioned in the previous section, for the model we are using, our scheme achieves higher rates than the known baseline scheme. However, such model is not realistic, rather, it is a worst-case scenario model. In this section, we compare our scheme against the message-wise baseline scheme. We first compare both without propagation delay, fixing the same rate $R$ and delay constraint $T$. Then, we compare both with propagation delay. In this case, we compare the two closest rates, above and below ours, as well as a suboptimal set of parameters for the message-wise which leads to the same rate under same rate under delay constraint.

We evaluate the different coding schemes over Gilbert-Elliott (GE) channels \cite{gilbert1960capacity,elliott1963estimates}, which is a well-known statistical channel that is useful for approximating packet losses experienced at the network layer \cite{hasslinger2008gilbert,hohlfeld2008packet}. The GE channel is a two-state Markov model that consists of a good state and a bad state. In the good state, each channel packet is lost with probability $\varepsilon\in [0,1)$, whereas, in the bad state, each channel packet is lost with probability 1. The average loss rate of the GE channel is given by 
\ifdefined\FULLVERSION
\begin{align*}
    \frac{\beta}{\beta+\alpha}\cdot\varepsilon+\frac{\beta}{\beta+\alpha}.
\end{align*}
\else
$\frac{\beta}{\beta+\alpha}\cdot\varepsilon+\frac{\beta}{\beta+\alpha}$,
\fi
where $\alpha$ and $\beta$ are the transition probabilities from the good state to the bad state and vice-versa.

As long as the channel stays in the bad state, the channel behaves as a burst erasure channel. In contrast, the channel behaves like i.i.d. erasure channel when the channel stays in a good state. We simulated GE channels with $(\alpha,\beta,\varepsilon)=(0.01,0.3,\varepsilon)$ where different random seed was used for each link ensuring there is no correlation between the erasures on each hop.

\begin{figure}
    \centering
    \begin{subfigure}{.45\textwidth}
        \centering
        \includegraphics[width=\textwidth, trim = .5cm 6.7cm 1.3cm 6.7cm, clip]{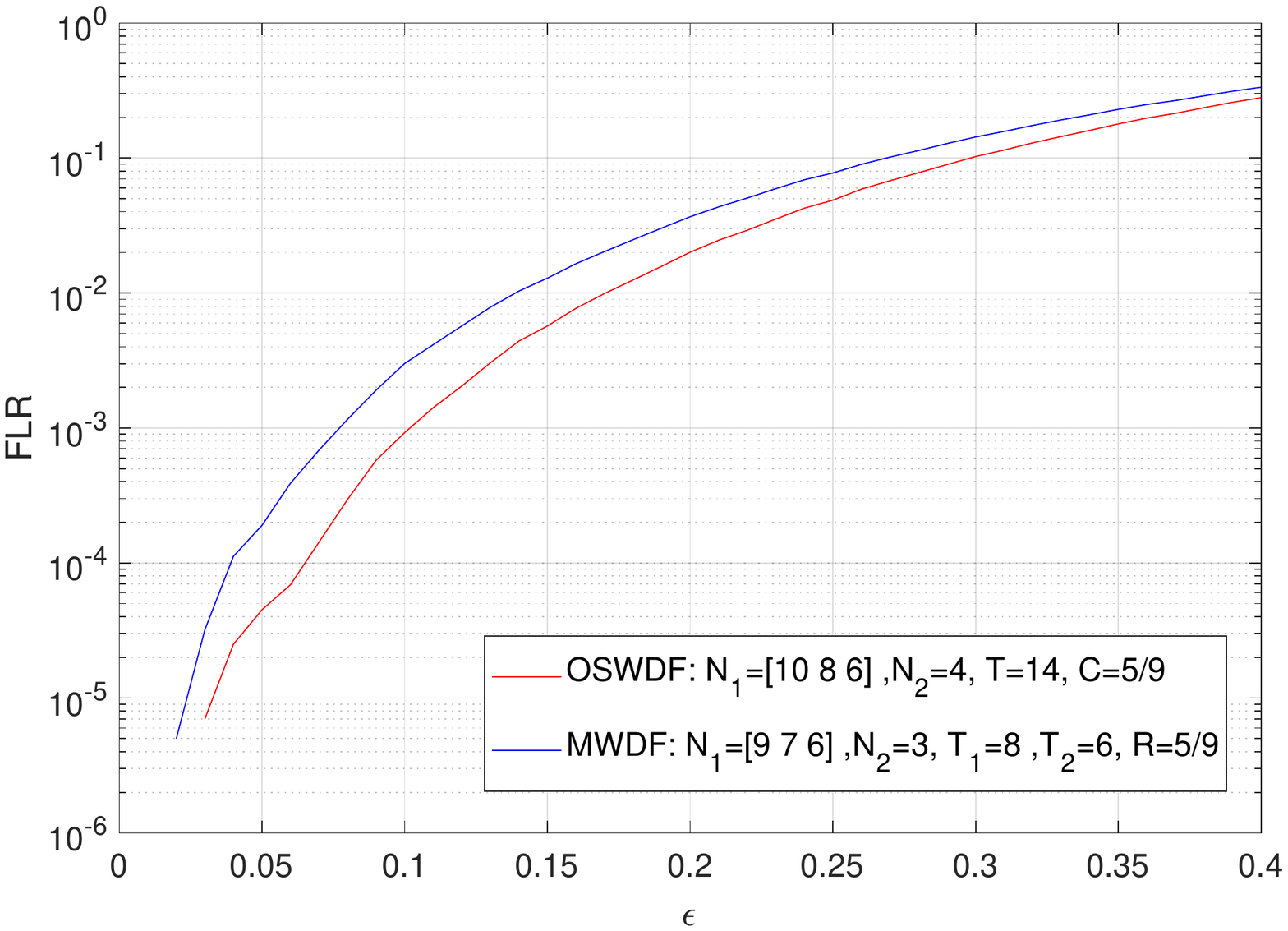}
        \caption{i.i.d. channel.}
        \label{fig:iidch}
    \end{subfigure}
    \begin{subfigure}{.45\textwidth}
    \centering
        \includegraphics[width=\textwidth, trim = .5cm 6.5cm 1.3cm 6.5cm, clip]{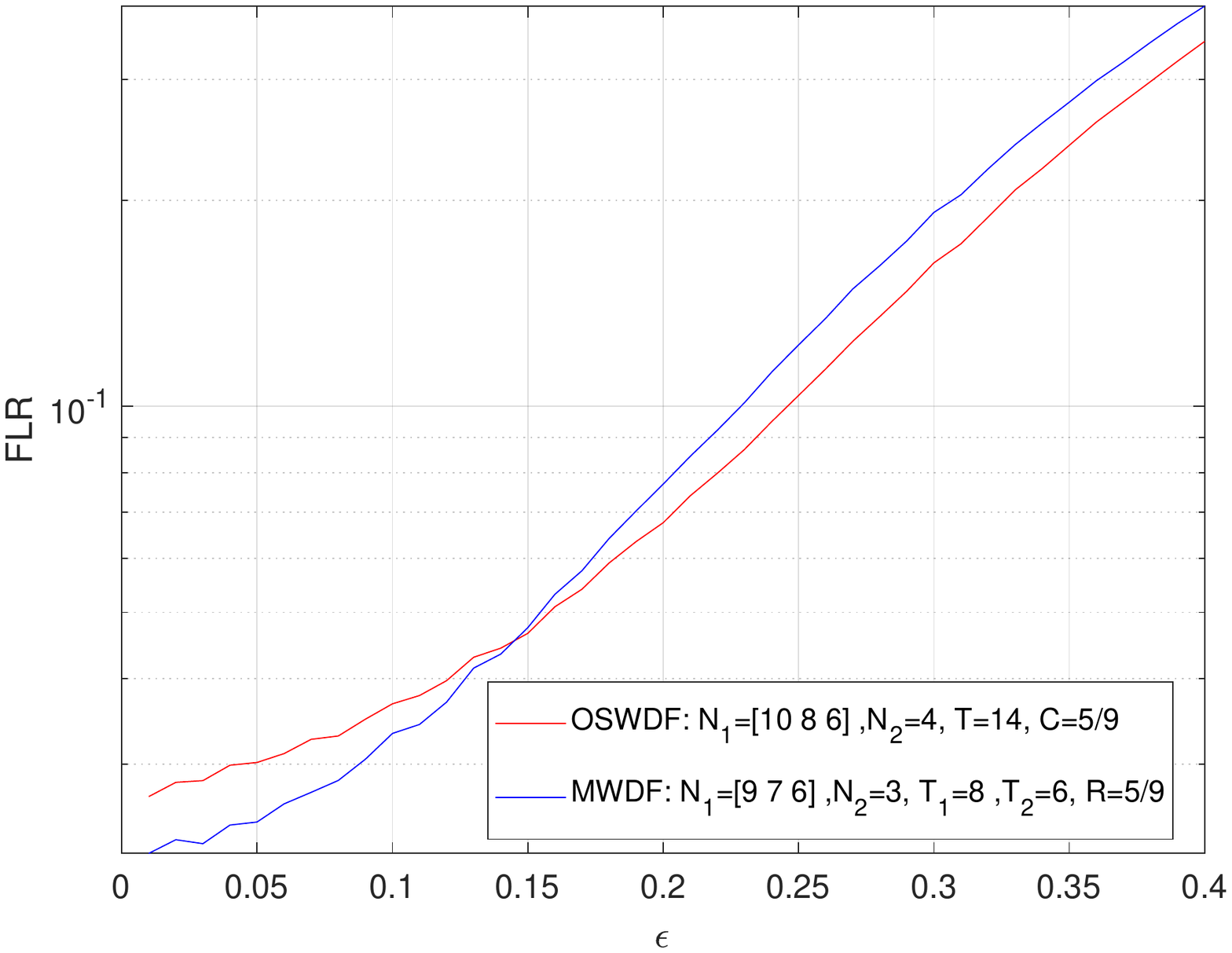}
        \caption{Gilbert-Elliott channel.}
        \label{fig:gech}
    \end{subfigure}
    \caption{Comparison between proposed scheme and message-wise baseline scheme without propagation delay.}\label{fig:nopropdelay}
\end{figure}

The results without propagation delay are shown in Fig.~\ref{fig:nopropdelay}. It can be seen that, for the i.i.d. channel, our scheme strictly outperforms the message-wise scheme. For the Gilbert-Elliott model, however, there is a region in which the proposed scheme is worse, but also a significant region where it is better. Empirical evidence suggests that the using systematic codes in the relay (in the traditional sense, that is, the source packets being transmitted as information symbols---not to be confused with the definition of ``systematic with respect to its own message'' used in our paper) is beneficial when handling burst erasures. That may explain why, for small $\epsilon$, the MWDF scheme is able to outperform our scheme under the Gilbert-Elliott model.

\begin{figure}
    \centering
    \begin{subfigure}{.45\textwidth}
        \centering
        \includegraphics[width=\textwidth, trim = .5cm 6.7cm 1.3cm 6.7cm, clip]{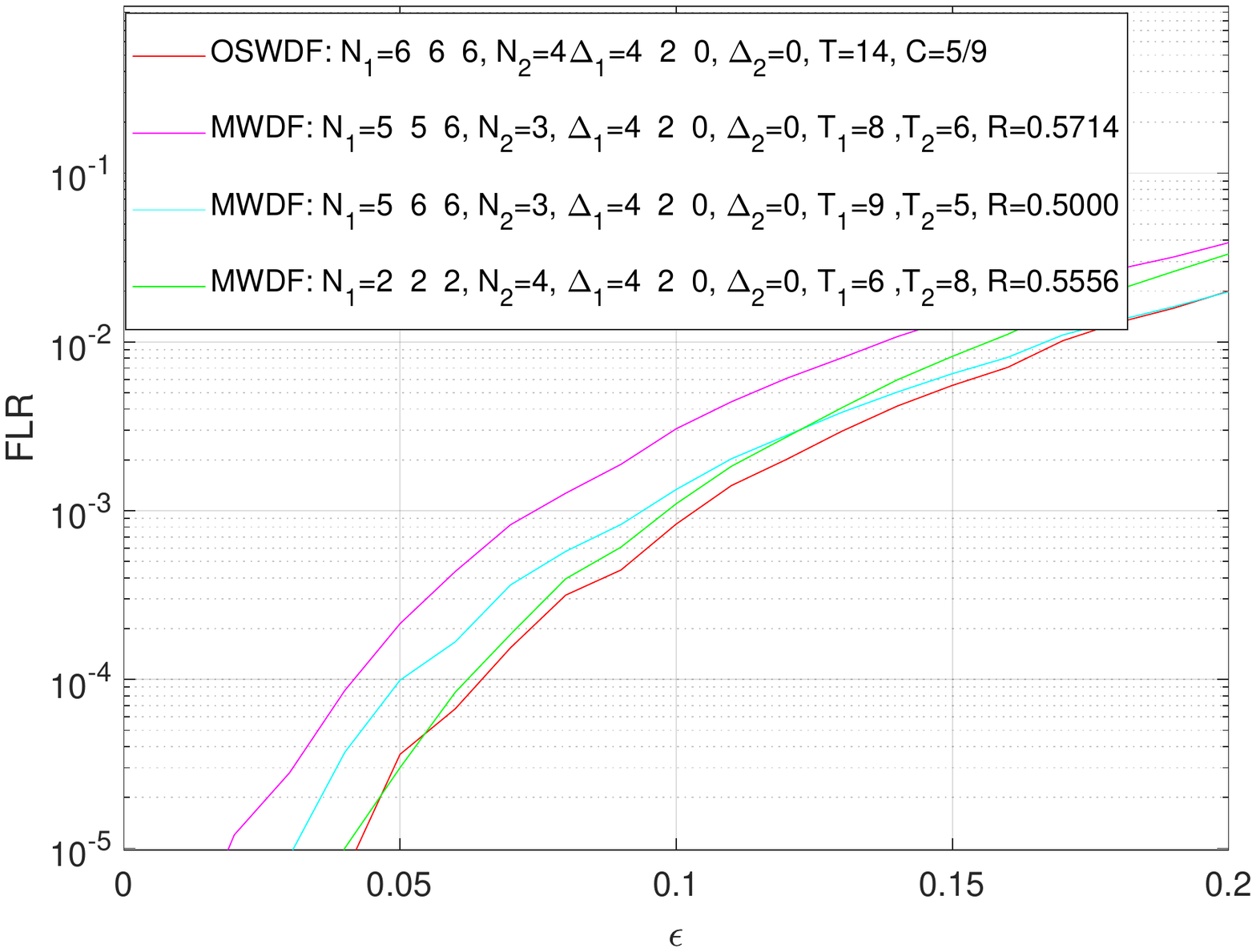}
        \caption{i.i.d. channel.}
        \label{fig:iidchprop}
    \end{subfigure}
    \begin{subfigure}{.45\textwidth}
    \centering
        \includegraphics[width=\textwidth, trim = .5cm 6.7cm 1.3cm 6.7cm, clip]{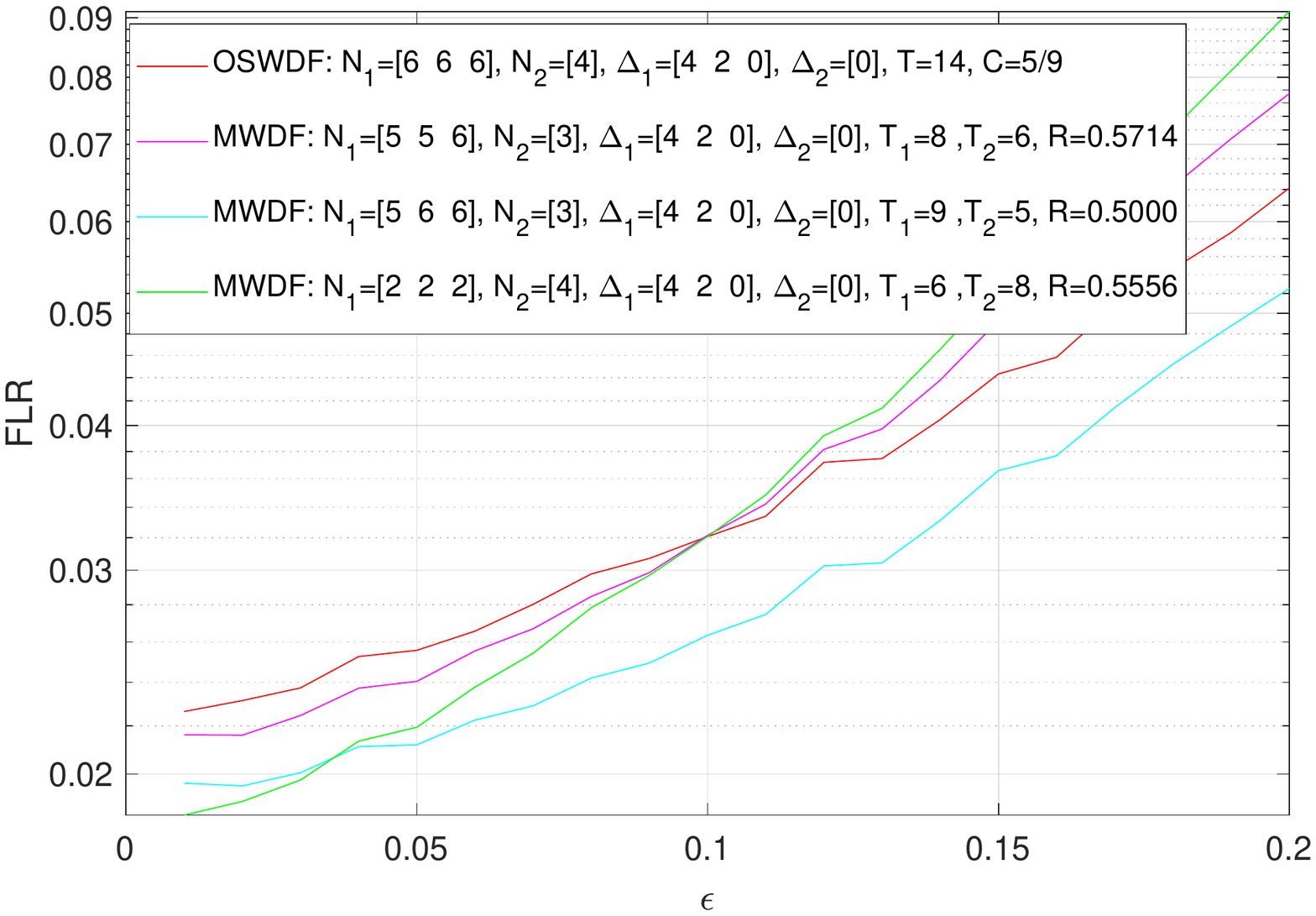}
        \caption{Gilbert-Elliott channel.}
        \label{fig:gechprop}
    \end{subfigure}
    \caption{Comparison between proposed scheme and message-wise baseline scheme with propagation delay.}\label{fig:propdelay}
\end{figure}

The results with propagation delay are shown in Fig.~\ref{fig:propdelay}. It can be seen that, again, under the i.i.d. model we are consistently better than the message-wise baseline scheme. As for the Gilbert-Elliott channel, there is still a significant region where we can outperform the message-wise scheme, except when the rate of the baseline scheme is considerably lower, in which case it achieves a lower packet loss rate than ours. However, this is at the cost of a significant rate loss, however, it is presented since it is the closest optimal rate just below our scheme.


\section{Conclusion}

We have generalized the three-node network topology described in \cite{Silas2019} to handle multiple links between each hop with different channel conditions. We have presented an analysis on the achievable rate under a maximal number of erasures model, including an upper bound which is achieved under certain scenarios, two baseline schemes which generalize the schemes previously proposed in \cite{Silas2019} and provide lower bounds on the achievable rate, and finally an optimization algorithm which can be used to design schemes that achieve, to the best of our knowledge, the highest rates known at the moment.

Further, we simulate codes designed using the optimized parameters under more realistic channel models and show that, for a considerable region of operation, our scheme is able to outperform the aforementioned baseline schemes. 

Our topology and proposed schemes have a broad range of applications in which they may be useful. It is also important to note that both the CSWDF scheme and our proposed OSWDF scheme employ separate coding strategy. For that reason, they can be easily adapted to suit a network containing multiple source nodes, such as a multi-user applications, and future research should explore it.

While the authors have not observed any improvement in the achievable rate using joint encoding rather than separate coding under the maximal number of erasures model, there is a noticeable improvement in the simulations, since one clean link may compensate for an excessive number of erasures in other links. For this reason, we believe studying symbol-wise joint encoding is an interesting research for future works.

\bibliographystyle{IEEEtran}
\bibliography{references}

\begin{thebibliography}{10}
\providecommand{\url}[1]{#1}
\csname url@samestyle\endcsname
\providecommand{\newblock}{\relax}
\providecommand{\bibinfo}[2]{#2}
\providecommand{\BIBentrySTDinterwordspacing}{\spaceskip=0pt\relax}
\providecommand{\BIBentryALTinterwordstretchfactor}{4}
\providecommand{\BIBentryALTinterwordspacing}{\spaceskip=\fontdimen2\font plus
\BIBentryALTinterwordstretchfactor\fontdimen3\font minus
  \fontdimen4\font\relax}
\providecommand{\BIBforeignlanguage}[2]{{%
\expandafter\ifx\csname l@#1\endcsname\relax
\typeout{** WARNING: IEEEtran.bst: No hyphenation pattern has been}%
\typeout{** loaded for the language `#1'. Using the pattern for}%
\typeout{** the default language instead.}%
\else
\language=\csname l@#1\endcsname
\fi
#2}}
\providecommand{\BIBdecl}{\relax}
\BIBdecl

\bibitem{Silas2019}
S.~L. {Fong}, A.~{Khisti}, B.~{Li}, W.~{Tan}, X.~{Zhu}, and
  J.~{Apostolopoulos}, ``Optimal streaming erasure codes over the three-node
  relay network,'' in \emph{2019 IEEE International Symposium on Information
  Theory (ISIT)}, July 2019, pp. 3077--3081.

\bibitem{GuaranteedMultihop2021}
E.~Domanovitz, G.~K. Facenda, A.~Khisti, W.-T. Tan, and J.~Apostolopoulos,
  ``Guaranteed rate of streaming erasure codes over multi-link multi-hop
  network,'' in \emph{2021 IEEE Information Theory Workshop (ITW)}, 2021, pp.
  1--6.

\bibitem{lin1984automatic}
S.~Lin, D.~J. Costello, and M.~J. Miller, ``Automatic-repeat-request
  error-control schemes,'' \emph{IEEE Communications magazine}, vol.~22,
  no.~12, pp. 5--17, 1984.

\bibitem{lockitt1975selective}
J.~Lockitt, A.~Gatfield, and T.~Dobyns, ``A selective repeat {ARQ} system,'' in
  \emph{dsc}, 1975, pp. 189--195.

\bibitem{weldon1982improved}
E.~Weldon, ``An improved selective-repeat {ARQ} strategy,'' \emph{IEEE
  Transactions on Communications}, vol.~30, no.~3, pp. 480--486, 1982.

\bibitem{comroe1984arq}
R.~Comroe and D.~Costello, ``{ARQ} schemes for data transmission in mobile
  radio systems,'' \emph{IEEE Journal on Selected Areas in Communications},
  vol.~2, no.~4, pp. 472--481, 1984.

\bibitem{Huang2010Skype}
T.~{Huang}, P.~{Huang}, K.~{Chen}, and P.~{Wang}, ``Could skype be more
  satisfying? a qoe-centric study of the fec mechanism in an internet-scale
  voip system,'' \emph{IEEE Network}, vol.~24, no.~2, pp. 42--48, 2010.

\bibitem{gallager1962low}
R.~Gallager, ``Low-density parity-check codes,'' \emph{IRE Transactions on
  information theory}, vol.~8, no.~1, pp. 21--28, 1962.

\bibitem{mackay1996near}
D.~J. MacKay and R.~M. Neal, ``Near shannon limit performance of low density
  parity check codes,'' \emph{Electronics letters}, vol.~32, no.~18, pp.
  1645--1646, 1996.

\bibitem{martinian2004burst}
E.~Martinian and C.-E. Sundberg, ``Burst erasure correction codes with low
  decoding delay,'' \emph{IEEE Transactions on Information theory}, vol.~50,
  no.~10, pp. 2494--2502, 2004.

\bibitem{leong2012erasure}
D.~Leong and T.~Ho, ``Erasure coding for real-time streaming,'' in \emph{2012
  IEEE International Symposium on Information Theory Proceedings}, 2012, pp.
  289--293.

\bibitem{badr2013streaming}
A.~Badr, A.~Khisti, W.-T. Tan, and J.~Apostolopoulos, ``Streaming codes for
  channels with burst and isolated erasures,'' in \emph{2013 Proceedings IEEE
  INFOCOM}.\hskip 1em plus 0.5em minus 0.4em\relax IEEE, 2013, pp. 2850--2858.

\bibitem{ho2003randomized}
T.~Ho, M.~Medard, J.~Shi, M.~Effros, and D.~R. Karger, ``On randomized network
  coding,'' in \emph{Proceedings of the Annual Allerton Conference on
  Communication Control and Computing}, vol.~41, no.~1.\hskip 1em plus 0.5em
  minus 0.4em\relax Citeseer, 2003, pp. 11--20.

\bibitem{JoshiWornell2012}
G.~{Joshi}, Y.~{Kochman}, and G.~W. {Wornell}, ``On playback delay in streaming
  communication,'' in \emph{2012 IEEE International Symposium on Information
  Theory Proceedings}, 2012, pp. 2856--2860.

\bibitem{Karzand2017}
M.~{Karzand}, D.~J. {Leith}, J.~{Cloud}, and M.~{Médard}, ``Design of {FEC}
  for low delay in {5G},'' \emph{IEEE Journal on Selected Areas in
  Communications}, vol.~35, no.~8, pp. 1783--1793, 2017.

\bibitem{badr2017layered}
A.~Badr, P.~Patil, A.~Khisti, W.-T. Tan, and J.~Apostolopoulos, ``Layered
  constructions for low-delay streaming codes,'' \emph{IEEE Transactions on
  Information Theory}, vol.~63, no.~1, pp. 111--141, 2017.

\bibitem{badr2017fec}
A.~Badr, A.~Khisti, W.-t. Tan, X.~Zhu, and J.~Apostolopoulos, ``{FEC} for
  {V}o{IP} using dual-delay streaming codes,'' in \emph{IEEE INFOCOM 2017-IEEE
  Conference on Computer Communications}.\hskip 1em plus 0.5em minus
  0.4em\relax IEEE, 2017, pp. 1--9.

\bibitem{Rashmi2018}
M.~{Rudow} and K.~V. {Rashmi}, ``Streaming codes for variable-size arrivals,''
  in \emph{2018 56th Annual Allerton Conference on Communication, Control, and
  Computing (Allerton)}, 2018, pp. 733--740.

\bibitem{krishnan2018rate}
M.~N. Krishnan and P.~V. Kumar, ``Rate-optimal streaming codes for channels
  with burst and isolated erasures,'' in \emph{2018 IEEE International
  Symposium on Information Theory (ISIT)}.\hskip 1em plus 0.5em minus
  0.4em\relax IEEE, 2018, pp. 1809--1813.

\bibitem{fong2019optimal}
S.~L. {Fong}, A.~{Khisti}, B.~{Li}, W.~{Tan}, X.~{Zhu}, and
  J.~{Apostolopoulos}, ``Optimal streaming codes for channels with burst and
  arbitrary erasures,'' \emph{IEEE Transactions on Information Theory},
  vol.~65, no.~7, pp. 4274--4292, 2019.

\bibitem{domanovitz2019explicit}
E.~Domanovitz, S.~L. Fong, and A.~Khisti, ``An explicit rate-optimal streaming
  code for channels with burst and arbitrary erasures,'' \emph{arXiv preprint
  arXiv:1904.06212}, 2019.

\bibitem{KrishnanLowField2020}
M.~N. {Krishnan}, D.~{Shukla}, and P.~V. {Kumar}, ``Low field-size,
  rate-optimal streaming codes for channels with burst and random erasures,''
  \emph{IEEE Transactions on Information Theory}, pp. 1--1, 2020.

\bibitem{domanovitz2020streaming}
\BIBentryALTinterwordspacing
E.~Domanovitz, A.~Khisti, W.-T. Tan, X.~Zhu, and J.~Apostolopoulos, ``Streaming
  erasure codes over multi-hop relay network,'' \emph{CoRR}, vol. 2006.05951,
  2020. [Online]. Available: \url{https://arxiv.org/abs/2006.05951}
\BIBentrySTDinterwordspacing

\bibitem{facenda2021streaming}
\BIBentryALTinterwordspacing
G.~K. Facenda, E.~Domanovitz, A.~Khisti, W.~Tan, and J.~G. Apostolopoulos,
  ``Streaming erasure codes over the multiple access relay channel,''
  \emph{CoRR}, vol. abs/2101.11117, 2021. [Online]. Available:
  \url{https://arxiv.org/abs/2101.11117}
\BIBentrySTDinterwordspacing

\bibitem{AdaptiveRelay}
M.~N. Krishnan, G.~K. Facenda, E.~Domanovitz, A.~Khisti, W.~{Tan}, and
  J.~{Apostolopoulos}, ``High rate streaming codes over the three-node relay
  network,'' in \emph{2021 IEEE Information Theory Workshop (ITW2021)}, October
  2021.

\bibitem{Cohen2019}
\BIBentryALTinterwordspacing
A.~Cohen, G.~Thiran, V.~B. Bracha, and M.~M{\'{e}}dard, ``Adaptive causal
  network coding with feedback for multipath multi-hop communications,''
  \emph{CoRR}, vol. abs/1910.13290, 2019. [Online]. Available:
  \url{http://arxiv.org/abs/1910.13290}
\BIBentrySTDinterwordspacing

\bibitem{Cohen2021}
A.~Cohen, G.~Thiran, V.~B. Bracha, and M.~Médard, ``Adaptive causal network
  coding with feedback for multipath multi-hop communications,'' \emph{IEEE
  Transactions on Communications}, vol.~69, no.~2, pp. 766--785, 2021.

\bibitem{Trestian2018heterogeneous}
R.~{Trestian}, I.~{Comsa}, and M.~F. {Tuysuz}, ``Seamless multimedia delivery
  within a heterogeneous wireless networks environment: Are we there yet?''
  \emph{IEEE Communications Surveys Tutorials}, vol.~20, no.~2, pp. 945--977,
  2018.

\bibitem{Hansen2015}
J.~{Hansen}, D.~E. {Lucani}, J.~{Krigslund}, M.~{Medard}, and F.~H.~P.
  {Fitzek}, ``Network coded software defined networking: enabling 5g
  transmission and storage networks,'' \emph{IEEE Communications Magazine},
  vol.~53, no.~9, pp. 100--107, 2015.

\bibitem{SaadatMultipath2018}
\BIBentryALTinterwordspacing
S.~Saadat, D.~Chen, and T.~Jiang, ``Multipath multihop mmwave backhaul in
  ultra-dense small-cell network,'' \emph{Digital Communications and Networks},
  vol.~4, no.~2, pp. 111--117, 2018. [Online]. Available:
  \url{https://www.sciencedirect.com/science/article/pii/S2352864817300317}
\BIBentrySTDinterwordspacing

\bibitem{Multiplex2018}
A.~{Badr}, D.~{Lui}, A.~{Khisti}, W.~{Tan}, X.~{Zhu}, and J.~{Apostolopoulos},
  ``Multiplexed coding for multiple streams with different decoding delays,''
  \emph{IEEE Transactions on Information Theory}, vol.~64, no.~6, pp.
  4365--4378, June 2018.

\bibitem{Fong2019multiplex}
S.~L. {Fong}, A.~{Khisti}, B.~{Li}, W.~{Tan}, X.~{Zhu}, and
  J.~{Apostolopoulos}, ``Optimal multiplexed erasure codes for streaming
  messages with different decoding delays,'' in \emph{2019 IEEE International
  Symposium on Information Theory (ISIT)}, 2019, pp. 3082--3086.

\bibitem{gilbert1960capacity}
E.~N. Gilbert, ``Capacity of a burst-noise channel,'' \emph{Bell system
  technical journal}, vol.~39, no.~5, pp. 1253--1265, 1960.

\bibitem{elliott1963estimates}
E.~O. Elliott, ``Estimates of error rates for codes on burst-noise channels,''
  \emph{The Bell System Technical Journal}, vol.~42, no.~5, pp. 1977--1997,
  1963.

\bibitem{hasslinger2008gilbert}
G.~Ha{\ss}linger and O.~Hohlfeld, ``The {G}ilbert-{E}lliott model for packet
  loss in real time services on the internet,'' in \emph{14th GI/ITG
  Conference-Measurement, Modelling and Evalutation of Computer and
  Communication Systems}.\hskip 1em plus 0.5em minus 0.4em\relax VDE, 2008, pp.
  1--15.

\bibitem{hohlfeld2008packet}
O.~Hohlfeld, R.~Geib, and G.~Ha{\ss}linger, ``Packet loss in real-time
  services: Markovian models generating {Q}o{E} impairments,'' in \emph{2008
  16th Interntional Workshop on Quality of Service}.\hskip 1em plus 0.5em minus
  0.4em\relax IEEE, 2008, pp. 239--248.

\end{thebibliography}
\newpage
\appendix

\section{Proofs}
\subsection{Proof of Lemma~\ref{lemma:df}}
\begin{IEEEproof}
	This follows directly from applying the decode-and-forward strategy to the relay. Let
	\begin{equation}
	\bs''_t = [\bs_{t - (T^{(1)}[1] )}[1], \bs_{t - (T^{(1)}[2])}[2], \ldots, \bs_{t - (T^{(1)}[k])}[k]].
	\end{equation}
	From assumption, the relay has access to the vector $\bs''_t$ at time $t$. The relay can then apply the permutation
	\begin{equation}
	\bs'_t = \pi \bs''_t
	\end{equation}
	and encode such sequence. Now, consider the $j$th symbol. It has been recovered with a delay $T^{(1)}[j]$ by the relay, then it has been permuted into the $j'$ symbol and recovered with delay $T^{(2)}[j']$ by the destination. For all $j$ and all permutations, the destination has recovered the symbol by time $T^{(1)}[j] + T^{(2)}[j']$, where $j'$ is defined by the permutation.
\end{IEEEproof}

\subsection{Proof of Lemma~\ref{lemma:conc}}
\begin{IEEEproof}
	Consider the concatenation code of both codes. Since the first code is able to achieve delay spectrum $\bT'$ under $\bN$ erasures, the first $k'$ symbols can be recovered certainly at times $\bT'$. Similarly, since the second code achieves delay spectrum $\bT''$ under $\bN$ erasures, the last $k''$ symbols can be recovered certainly at times $\bT''$. Thus, the concatenation code achieves the delay spectrum $[\bT', \bT'']$ under $\bN$ erasures.
\end{IEEEproof}

\subsection{Proof of Lemma~\ref{lemma:permutation}}
\begin{IEEEproof}
    This follows directly from the choice of encoder. Let us denote by $f_{t}(\{\bs_j\}_{j=0}^{t}$ the encoding function of a code that achieves delay spectrum $\bT$ under $N$ erasures. Without loss of generality, let us analyze the first two source symbols. This code is able to recover them by times $T[1]$ and $T[2]$, respectively. Now, consider the code generated by applying the function to $\bs'$, i.e., $f_{t}(\{ \bs'_j\}_{j=0}^{t})$, where $\bs' = \pi \bs$. Again without loss of generality, assume the permutation swaps the first two positions. Then, $\bs'[1]$ is now recovered at time $T[1]$, and $\bs'[2]$ is recovered at time $T[2]$. However, $\bs'[1] = \bs[2]$, therefore, $\bs[2]$ is recovered at time $T[1]$, and similarly for $\bs[1]$.
    
    It is easy to see that, in general, applying the desired permutation over the source symbols before encoding results in the desired permuted delay spectrum. Thus, if a code is able to achieve delay spectrum $\bT$ under $N$ erasures, permuting the source symbols and using the same code suffices to acheive delay spectrum $\pi \bT$ under $N$ erasures.
\end{IEEEproof}

\subsection{Proof of Lemma~\ref{lemma:lb}}
\begin{IEEEproof}
    First, recall that, in order to a delay spectrum to be achievable under $N$ erasures, the decoder must be able to recover the $k^{(g)}[j]$ information symbols at time $T^{(g)}[j]$ with no ambiguity as long as the number of erasures is at most $N$.
    
	In order to prove the lemma, we make a counting argument similar to \cite{Multiplex2018} and \cite{Silas2019}. The argument can be formalized in terms of entropy, as seen in the mentioned papers. Consider $N$ erasures at the first $N$ positions, and other $N$ erasures after time $T^{(g)}[j]$, as in Fig.~\ref{erpat}.
	
	\begin{figure}[h!]
    \centering
    \includegraphics[]{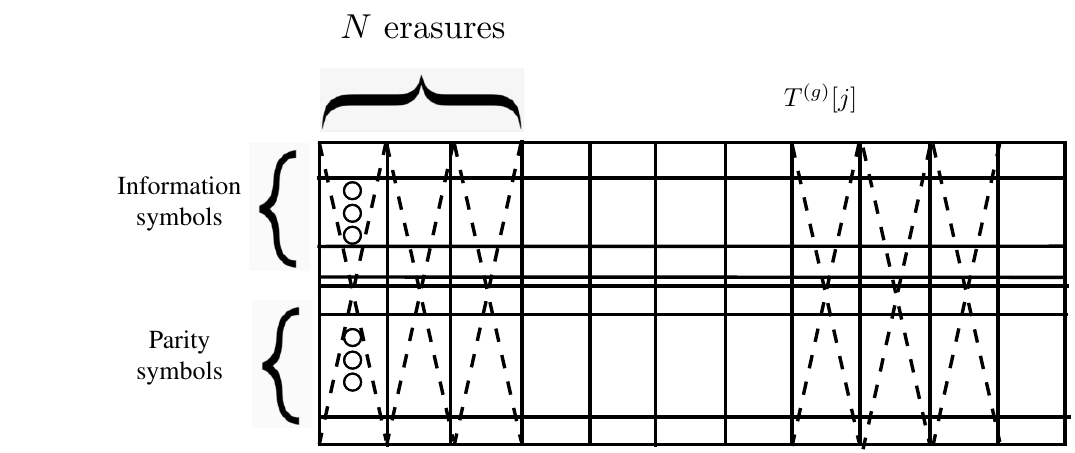}
    \caption{Periodic erasure pattern.} \label{erpat}
    \end{figure}
	
	More precisely, due to the systematic assumption, we have the following condition
	\begin{align}
	    H(\bs_{N:T^{(g)}[j]} | \bx_{N : T^{(g)}[j]}) = 0
	\end{align}
	that is, the source packets at non-erased times must be fully recoverable.
	
	Furthermore, since all symbols with delays smaller than or equal to $T^{(g)}[j]$ must be recoverable with that delay, we also must have that
	\begin{align}
	    H(\bs_{0: N - 1} | \bx_{N : T^{(g)}[j]}, \bs_{0:N-1}[1:j-1]) = 0 
	\end{align}
	that is, if we are given the symbols with higher delay for the erased packets, then we must be able to fully recover the erased packets using only the available information from $\bx$ up to time $T^{(g)}[j]$. 
	
	Then, we can write
	\begin{align}
	    \sum_{t = N}^{T^{(g)}[j]} H(\bx_t) + \sum_{t = 0}^{N-1} H(\bs_t[1:j-1]) &\geq H(\bx_{N : T^{(g)}}, \bs_{0:N-1}[1:j-1]) \\
	    &= H(\bs_{0:T^{(g)}}) \\ &= \sum_{t = 0}^{T^{(g)}} H(\bs_t).
	\end{align}
	Finally, by noting that $H(\bs_t) = k$ and $H(\bx_t) = n$, and further that $k = \sum_{j = 1}^{\ell^{(g)}} k^{(g)}[j]$, we can write 
	\begin{equation}
	(T^{(g)}[j] + 1 - N)n \geq (T^{(g)}[j] + 1) \sum_{\ell=j}^{\ell^{(g)}} k^{(g)}[\ell] + (T^{(g)}[j] + 1 - N) \sum_{\ell=1}^{j-1} k^{(g)}[\ell] .\label{ConverseMultiplex}
	\end{equation}
	
	If this condition does not hold for some $T^{(g)}[j]$, there is ambiguity in the recovery of the symbols and the information of the $k^{(g)}[j]$ symbols can not be recovered by the deadline, therefore, this delay spectrum is not achievable under $N$ erasures.
	
	This condition can then be rewritten as
	\begin{align}
	T^{(g)}[j] \geq N \frac{\left(1 - \sum_{\ell = 1}^{j-1} \frac{k^{(g)}[\ell]}{n} \right)}{\left(1 - \sum_{\ell=1}^{\ell^{(g)}}\frac{k^{(g)}[\ell]}{n}\right)} - 1. \label{eq:MaxNumberOfSymbols}
	\end{align}
	
	Now, recall that $k = \sum_{\ell=1}^{\ell^{(g)}} k^{(g)}[\ell]$, then we have
	\begin{align}
	T^{(g)}[j] \geq \frac{N n}{n - k}\left(1 - \sum_{\ell = 1}^{j-1} \frac{k^{(g)}[\ell]}{n} \right) - 1.
	\end{align}
	
	The implicit assumption in the erasure pattern and the deduction is that $T^{(g)}[j] \geq T^{(g)}[j + 1]$ and $T^{(g)}[j + 1] + N \geq T^{(g)}[j]$. However, we can always set $T^{(g)}[j+1] = T^{(g)} - 1$, and simply have some $k^{g}[j] = 0$, thus such assumption is not a problem.
\end{IEEEproof}

\subsection{Proof of Corollary~\ref{corollary:bounds}}
\begin{IEEEproof}
	This follows immediately from rearranging Lemma~\ref{lemma:lb} and solving it for $T^{(g)}[2] = T^{(g)}[1] - 1$ and $T^{(g)}[\ell^{(g)}] = N$, with $\sum_{\ell = 1}^{\ell^{(g)}} k^{(g)}[\ell] = k - k^{(g)}[\ell^{(g)}]$.
\end{IEEEproof}
\begin{IEEEproof}[Proof of Corollary~\ref{corollary:recursion}]
    First, recall that $\sum_{\ell = 1}^{\ell^{(g)}} k^{(g)}[\ell] = k$. Then, let us rearrange \eqref{eq:DelayConLowBound} as follows
    \begin{align}
        \bT^{(g)}[j] &\geq \frac{N n}{n - k}\left(1 - \sum_{\ell = 1}^{j-1} \frac{k^{(g)}[\ell]}{n} \right) - 1 \\
        (\bT^{(g)}[j] + 1) \frac{n - k}{N} &\geq n - k + \sum_{\ell = j}^{\ell^{(g)}} k^{(g)}[\ell] \\
        \sum_{\ell = j}^{\ell^{(g)}} k^{(g)}[\ell] &\leq (\bT^{(g)}[j] + 1 - N) \frac{n - k}{N}.
    \end{align}
    Now, let us solve this for $j = \ell^{(g)}, \ell^{(g)} - 1, \ldots, 1$. We get
    \begin{align}
        k^{(g)}[\ell^{(g)}] &\leq \frac{n - k}{N} \\ 
        k^{(g)}[\ell^{(g)}] + k^{(g)}[\ell^{(g)}-1] &\leq 2 \frac{n - k}{N} \\
        &\vdots \\
        \sum_{\ell = 1}^{\ell^{(g)}} k^{(g)}[\ell] &\leq \ell^{(g)} \frac{n - k}{N}. \label{eq:finalcond}
    \end{align}
    Now, note that there is a maximum number of symbols that can be transmitted with the best delay, that is, there is a direct upper bound on $k^{(g)}[\ell^{(g)}]$, which is transmitted with the lowest delay (i.e., $N$). Furthermore, note that transmitting less symbols with the lowest delay only allows exactly that same number of symbols (i.e., the difference between the bound and the number of symbols transmitted) to be transmitted with worse delays, therefore, it is optimal to transmit with equality, as the only choice is to transmit those symbols with a better delay or not, and there is no reason to transmit with a worse delay. The equality for $k^{(g)}[1]$ comes from noting that $\ell^{(g)} = T^{(g)}[1] + 1 - N$ and $k^{(g)}[1] = k - \sum_{\ell=2}^{\ell^{(g)}} k[\ell]$, thus 
    \begin{align}
        k^{(g)}[1] &= k - \sum_{\ell=2}^{\ell^{(g)}} \\ 
        &= k - (\ell^{(g)} - 1) \frac{n - k}{N} \\
        &= k - (T^{(g)}[1] - N) \frac{n - k}{N} \\
        &= n - \frac{T^{(g)}[1]}{N}(n - k).
    \end{align}
\end{IEEEproof}

\subsection{Proof of Corollary~\ref{corollary:maxsym}}
\begin{IEEEproof}
	The proof follows directly from $\sum_{\ell = 1}^{j-1} k^{(g)}[\ell] \leq \sum_{\ell = 1}^{j-1} k^{\textrm{con}}[\ell]$, i.e.,
	\begin{align}
	T^{(g)}[j] &\geq \frac{n \cdot N}{n - k} \left(1 - \sum_{\ell = 1}^{j-1} \frac{k^{(g)}[\ell]}{n}\right)\\
	&\geq \frac{n \cdot N}{n - k} \left(1 - \sum_{\ell = 1}^{j-1} \frac{k^{\textrm{con}}[\ell]}{n}\right)
	\end{align}
	and rearranging the expression to isolate $k$. Finally, note that this must be true for all $j$, completing the proof.
\end{IEEEproof}

\subsection{Proof of Lemma~\ref{lemma:achiev}}
\begin{IEEEproof}
    Recall that we wish to construct an $(n, k)$ code such that, under $N$ erasures, $k^{(g)}[j]$ symbols are recovered at time $T^{(g)}[j]$ and
    \begin{equation}
        k^{(g)}[j] = \begin{cases}
        n - \frac{T^{(g)}[1]}{N}(n - k), &j = 1\\
        \frac{n - k}{N}, &j \geq 2
        \end{cases}.
    \end{equation}
    For the remaining of the proof, we denote $k^{(g)}[j] = \frac{n - k}{N}$, i.e., we assume $j \geq 2$. 

	Consider the following coding scheme: denote by $m = (T^{(g)}[1] + 1 - N)$ the number of possible delays. We concatenate $k^{(g)}[1] = n - \frac{T^{(g)}[1]}{N}(n - k)$ diagonal interleaving MDS codes with parameters $(N + m, m)$ and  $k^{(g)}[j] - k^{(g)}[1] =  \frac{T^{(g)}[1] + 1}{N}(n - k) - n$ interleaving MDS codes with parameters $(N + m - 1, m - 1)$.
	
	It follows directly from Lemma~\ref{lemma:intMDS} and Lemma~\ref{lemma:conc} that $k^{(g)}[1]$ symbols are recovered with delay $T^{(g)}[1]$ and that $k^{(g)}[j]$ symbols are recovered with every other delay (i.e., $T^{(g)}[j]$). From definition, we have $k = k^{(g)}[1] + \sum_{j=2}^{m} k^{(g)}[j]$, since the number of possible delays is exactly $m$.
	
	Therefore, it remains to show that the number of channel uses this proposed code uses, which we will briefly denote as $n'$, is the designed number of channel uses $n$.
	
	The number of channel uses of this concatenation is, from definition of concatenation, given by
	\begin{align}
	n'= &k^{(g)}[1] (N + m) + (N + m - 1)(k^{(g)}[j] - k^{(g)}[1]) \\
	&= (N + m - 1)k^{(g)}[j] + k^{(g)}[1] \overset{(b)}{=} N k^{(g)}[j] + k \\
	&\overset{(c)}{=} (n - k) + k = n
	\end{align}
	where $(a)$ follows from $k^{(g)}[1] + (m - 1)k^{(g)}[j] = k$ and $(b)$ follows from $k^{(g)}[j] = \frac{1}{N}(n - k)$.
	
	Now, note that this requires $n - \frac{T[1]}{N}(n - k)$ and $\frac{T[1] + 1}{N} (n - k)$ to be integer. Both are true from assumption. This is intuitive from the construction: we are concatenating codes of the form $(N + k', k')$, that is, all the codes that composite the code have exactly $N$ parity symbols, thus the number of parity symbols has to be some multiple of $N$. 
\end{IEEEproof}

\subsection{Proof of Lemma~\ref{lemma:sepcod}}
\begin{IEEEproof}
	Consider a code following the separate coding strategy using each delay-spectrum-achieving single-link code for each respective link. It is clear that the first $k_1$ symbols are recoverable at times $\bT_1$, then the next $k_2$ symbols are recoverable at times $\bT_2$, etc., from assumption that the code used in each link is able to recover those symbols at those times. Then, the achievable delay spectrum is simply the concatenation of all delay spectra.
\end{IEEEproof}


\subsection{Proof of Proposition~\ref{prop:concSWrate}}
\begin{IEEEproof}
    Let us focus the analysis on the single-link network resulting from connecting the $i$th link in the first hop with the $j$th link in the second hop. Let us transmit using a streaming code for the single link three-node network presented in \cite{Silas2019} with the parameters
    \begin{align}
        k_{i, j} &= (T + 1 - N^{(1)}_i - N^{(2)}_j)^+ \\
        n^{(1)}_{i} &= T + 1 - N^{(2)}_j \\
        n^{(2)}_{j} &= T + 1 - N^{(1)}_i
    \end{align}
    we know, from the results in \cite{Silas2019}, that this code is able to recover these $k_{i, j}$ symbols of information at time $T$. Now, for the $i$th link in the first hop, we repeat this analysis for all $j \in \{ 1, \ldots, L_{r, d} \}$, and concatenate all of these codes together. That way, using the results from Lemma~\ref{lemma:conc}, we get
    \begin{align}
        k^{(1)}_i &= \sum_{j = 1}^{L_{r,d}} (T + 1 - N^{(1)}_i - N^{(2)}_j)^+\\
        n^{(1)} &= n^{(1)}_i = \sum_{j = 1}^{L_{r,d}} T + 1 - N^{(2)}_j.
    \end{align}
    Again, from the parameters we choose and the results from \cite{Silas2019}, we know all the streaming codes used are capable of recovering the information symbols by the deadline.
    
    On the other hand, if we do the same for the $j$th link, concatenating all the codes formed by ``connecting'' the $j$th link to the $i$th link, $i \in \{1, \ldots, L_{s,r} \}$, we get
    \begin{align}
        k^{(2)}_j &= \sum_{i = 1}^{L_{s,r}} (T + 1 - N^{(2)}_j - N^{(1)}_i)^+\\
        n^{(2)} &= n^{(2)}_j = \sum_{i = 1}^{L_{s,r}} T + 1 - N^{(1)}_i.
    \end{align}
    
    Now note that, summing across all links, we have
    \begin{align}
        k &= \sum_{i=1}^{L_{s,r}} k^{(1)}_i = \sum_{j=1}^{L_{r,d}} k^{(2)}_j\\
        &= \sum_{i=1}^{L_{s,r}}\sum_{j=1}^{L_{r,d}} (T + 1 - N^{(1)}_i - N^{(2)}_j)^+.
    \end{align}
    
    Thus, the rate the scheme, as described, achieves, is given by
    \begin{align}
        R_{\textrm{con-SW}} = \frac{\sum_{i=1}^{L_{s,r}}\sum_{j=1}^{L_{r,d}} (T + 1 - N^{(1)}_i - N^{(2)}_j)^+}{\max(\sum_{j = 1}^{L_{r,d}} T + 1 - N^{(2)}_j, \sum_{i = 1}^{L_{s,r}} T + 1 - N^{(1)}_i )}.
    \end{align}
    
    Recalling that $\bar{N}^{(1)}$ is the average of the number of erasures in the first hop and $\bar{N}^{(2)}$ is the average of the number of erasures in the second hop, we get
    \begin{align}
        R_{\textrm{con-SW}} &= \frac{\sum_{i=1}^{L_{s,r}}\sum_{j=1}^{L_{r,d}} (T + 1 - N^{(1)}_i - N^{(2)}_j)^+}{\max(\sum_{j = 1}^{L_{r,d}} T + 1 - N^{(2)}_j, \sum_{i = 1}^{L_{s,r}} T + 1 - N^{(1)}_i )} \\
        &\geq \frac{\sum_{i=1}^{L_{s,r}}\sum_{j=1}^{L_{r,d}} (T + 1 - N^{(1)}_i - N^{(2)}_j)}{\max(\sum_{j = 1}^{L_{r,d}} T + 1 - N^{(2)}_j, \sum_{i = 1}^{L_{s,r}} T + 1 - N^{(1)}_i )}\\
        &=\frac{L_{s,r}\cdot L_{r,d}(T+1)-L_{s,r}\cdot\sum_{i=1}^{L_{r,d}}N_i^{(2)} - L_{r,d}\cdot\sum_{i=1}^{L_{s,r}}N_i^{(1)}}{\max\left(L_{r,d}(T+1)-\sum_{i=1}^{L_{r,d}}N_i^{(2)},L_{s,r}(T+1)-\sum_{i=1}^{L_{s,r}}N_i^{(1)}\right)}\\
        &= \frac{T+1-\bar{N}^{(2)}-\bar{N}^{(1)}}{\max\left(\frac{1}{L_{s,r}}\left(T-\bar{N}^{(2)}+1\right),\frac{1}{L_{r,d}}\left(T-\bar{N}^{(1)}+1\right)\right)}
    \end{align}
    which is the expression given in the proposition. Note that, in \cite{Silas2019}, the $(\cdot)^+$ operation is not necessary due to the assumption of $T \geq N^{(1)} + N^{(2)}$. In our paper, it is possible that, for some combination of links, we have $T < N^{(1)}_i + N^{(2)}_j$, and the overall rate is still greater than zero.
\end{IEEEproof}

\subsection{Proof of Proposition~\ref{prop:concSWoptimal}}
\begin{IEEEproof}
    The proof of this proposition follows easily from comparing \eqref{eq:explicit_R_con_SW} to the upper bound \eqref{eq:explicit_upper}. Note that the minimums (i.e., $N^{(1)}_{\textrm{min}}$ and $N^{(2)}_{\textrm{min}}$) have been replaced by the average (i.e., $\bar{N} {(1)}$ and $\bar{N}^{(2)}$). Therefore, it is easy to see that the rate this scheme achieves is equal to the upper bound when, in the bottleneck hop, the minimum number of erasures (of the other hop) is equal to the average number of erasures (of the other hop), i.e., when all links in the hop that is not the bottleneck are equal. Assume $\bar{N}^{(h)} > N^{(h)}_{\textrm{min}}$ for both hops. Then, note that $\frac{T + 1 - \bar{N}^{(h)} - N^{(h')}_i}{T + 1 - \bar{N}^{(h)}} < 1$, thus we have $\frac{T + 1 - \bar{N}^{(h)} - N^{(h')}_i}{T + 1 - \bar{N}^{(h)}} < \frac{T + 1 - N_{\textrm{min}}^{(h)} - N^{(h')}_i}{T + 1 - N_{\textrm{min}}^{(h)}}$, that is, the achieved rate is strictly less than the upper bound if the number of erasures is different.
\end{IEEEproof}

\subsection{Proof of Theorem~\ref{theorem:achievability}}
\begin{IEEEproof}
	This follows directly from our lemma and the choice of parameters. We now describe in detail the codes used.
	
	Again, without loss of generality, assume the first link is the bottleneck. Then, with the choice of each $k_i^{(1)}$, there exists a code that achieves the delay spectrum $\bG^{(1)}_i$. This follows from Lemma~\ref{lemma:achiev}. Then, by employing a separate code strategy, there exists a code that achieves the following delay spectrum
	\begin{align}
	\bG^{(1)} = \begin{bmatrix}
	(T - N^{(2)}_{\textrm{min}}, k^{(1)}[1])\\
	(T - N^{(2)}_{\textrm{min}} - 1, k^{(1)}[2])\\
	\vdots\\
	(N_{\textrm{min}}^{(1)}, k^{(1)}[\ell{^(1, g)}])
	\end{bmatrix}
	\end{align}
	which follows from Lemma~\ref{lemma:sepcod}. Then, we wish to pair these delays with delays in the second hop in such a way that the overall delay is equal to $T$. By applying a symbol-wise decode-and-forward strategy, the delay of each symbol is given by the sum of the delays in each hop for that symbol, therefore, the number of symbols that can be transmitted with a delay $T - T'$ in the second hop is the same number of symbols transmitted with delay $T'$ in the first hop. Conversely, if we transmit $k'$ symbols with delay $T'$ in the first hop and $k'$ symbols with delay $T - T'$ in the second hop, then all $k'$ symbols are recoverable in the destination at time $T$, as shown in Lemma~\ref{lemma:df}. Thus, it follows directly from Lemma~\ref{lemma:df} that, if we are able to satisfy the delay spectrum $\bG^{\textrm{con}}$, all symbols are recoverable by time $T$.
	
	The fact that our code design satisfies the constraint is a direct consequence of the choice of $k^{(2)}_i$. As shown by Corollary~\ref{corollary:maxsym}, the constraint on how many symbols can be transmitted at each delay can be translated into a constraint to the maximum number of symbols. Furthermore, since Corollary~\ref{corollary:maxsym} is a direct consequence of Lemma~\ref{lemma:lb}, the bound on Lemma~\ref{lemma:lb} being achievable implies the bound on Corollary~\ref{corollary:maxsym} being achievable. Thus, our code certainly satisfies the constraint. However, in order to design such a code, we may need to update the previously found parameters in order to satisfy the multiplicity constraint. However, Lemma~\ref{lemma:conc} shows that we may simply multiply the number of symbols for each delay by the same constant $N^{(2)}_i$, which is achievable by a concatenation of $N^{(2)}_i$ equal codes.
	
	Finally, it remains to justify the existence of negative entries in $\bk^{\textrm{con}}$ and our update rule. Since our code is proven to satisfy the constraint, such negative entries may only happen if a code that can recover the symbols earlier than the constraint exists. Our update rule can be translated as buffering (i.e., introducing an artificial delay) to these symbols, so they always arrive at time $T$ at the destination. Note that this is not required and we may allow the destination to recover some symbols earlier.
	
	Summarizing the proof: the existence of single-link codes that achieve each delay spectrum described is assured by Lemma~\ref{lemma:achiev}. The overall delay spectrum of the $L$-links code used being achievable is a consequence of Lemma~\ref{lemma:sepcod}. And finally, the overall delay of each symbol being less than or equal to $T$ follows from Lemma~\ref{lemma:df}, Corollary~\ref{corollary:maxsym} and Lemma~\ref{lemma:achiev}. 
\end{IEEEproof}

\section{Examples}

\subsection{Example~1}
\label{ex:singlelink}

\begin{figure}[h!]
\centering
\includegraphics[]{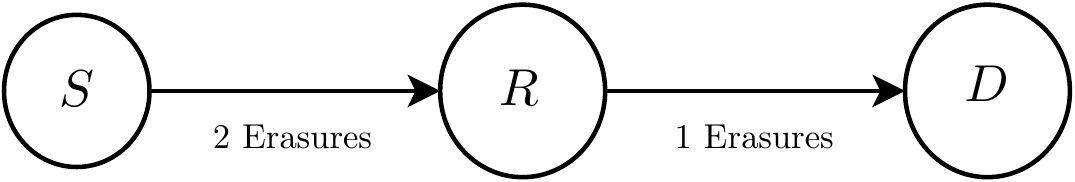}
\caption{Packets generated in the single-link relay network at time $i$ with $N_1=2$, $N_2=1$ and $T=5$.}
\end{figure}
    
    In order to achieve the upper bound for the upper links in Fig.~\ref{fig:achievex}, we use a $(5, 3)$ diagonal interleaving maximum distance separable (MDS) code, which is presented in Table~\ref{tab:sourcerelay1}, in the source to relay link. Note that, for any 2-erasure pattern, $\bs_t[3]$ can be recovered by time $t+2$, $\bs_t[2]$ can be recovered by time $t + 3$ and $\bs_t[1]$ can be recovered by time $t + 4$. Therefore, this code achieves delay spectrum $\bT = [4, 3, 2]$ under 2 erasures.

    \begin{table*}
    	\caption{Diagonal Interleaving $(5,3)$ MDS code.} \label{tab:sourcerelay1}
    	{\footnotesize
    	\begin{tabular}{|c|c|c|c|c|c|c|}
    		\hline
    		\backslashbox{Symbol}{Time} & $t$        & $t+1$          & $t+2$                       & $t+3$                                      & $t+4$                                        & $t+5$                                          \\ \hline
    		$\bx[1]$                                              & $\bs_t[1]$ & $\bs_{t+1}[1]$ & $\bs_{t+2}[1]$              & $\bs_{t+3}[1]$                             & $\bs_{t+4}[1]$                               & $\bs_{t+5}[1]$                                 \\ \hline
    		$\bx[2]$                                              & $\bs_t[2]$ & $\bs_{t+1}[2]$ & $\bs_{t+2}[2]$              & $\bs_{t+3}[2]$                             & $\bs_{t+4}[2]$                               & $\bs_{t+5}[2]$                                 \\ \hline
    		$\bx[3]$                                              & $\bs_t[3]$ & $\bs_{t+1}[3]$ & $\bs_{t+2}[3]$              & $\bs_{t+3}[3]$                             & $\bs_{t+4}[3]$                               & $\bs_{t+5}[3]$                                 \\ \hline
    		$\bx[4]$                                              &            &  $\bs_t[3]$     & \begin{tabular}{c}$\bs_{t+1}[3]$  \\ $+\bs_t[2]$\end{tabular} & \begin{tabular}{c}$\bs_{t+2}[3] + \bs_{t+1}[2]$ \\ $+\bs_{t}[1]$\end{tabular} & \begin{tabular}{c}$\bs_{t+3}[3] + \bs_{t+2}[2]$  \\ $+\bs_{t+1}[1]$\end{tabular} & \begin{tabular}{c}$\bs_{t+4}[3] + \bs_{t+3}[2]$  \\ $+\bs_{t+2}[1]$\end{tabular} \\ \hline
     		$\bx[5]$                                              &            &                & $\bs_t[3]$                  & $\bs_{t+1}[3] + 2\bs_t[2]$               & \begin{tabular}{c} $\bs_{t+2}[3] + 2\bs_{t+1}[2]$ \\ $+3\bs_{t}[1]$
     		\end{tabular} 
     		& \begin{tabular}{c}$\bs_{t+3}[3] + 2\bs_{t+2}[2]$  \\ $+3\bs_{t+1}[1]$\end{tabular} 
    		\\ \hline
    	\end{tabular}
    	}
    \end{table*}

From relay to destination, we use a $(4, 3)$ code, which is presented in Table~\ref{tab:relaydest1}. In this case, note that $\bs'_t[1]$ is recovered at time $t + 3$, $\bs'_t[2]$ is recovered at time $t+2$ and $\bs'_t[3]$ is recovered at time $t+1$, since there is only one erasure in the channel. Therefore, under 1 erasure, this code achieves a delay spectrum $\bT = [3, 2, 1]$. Note that we are able to set
\begin{align}
\bs'_t[1] = \bs_{t - 2}[3],\quad \bs'_t[2] = \bs_{t - 3}[2], \quad \bs'_t[3] = \bs_{t - 4}[1] \label{eq:matchdelays}
\end{align}
and, if we do so, the effective delay spectrum of the decode-and-forward scheme presented is $\bT = [4 + 1, 3 + 2, 2 + 3]$, and all source symbols are recovered at time $t + 5$. This can be seen in Table~\ref{tab:relaydestbefore}. We are allowed to do so because these symbols are recoverable by the relay at these time slots, due to the code used in the source to relay link.

\begin{table*}
    \centering
    \caption{Code used by the relay.} \label{tab:relaydest1}
    \begin{subtable}[]{\textwidth}
	\caption{Diagonal Interleaving $(4, 3)$ MDS code before relabeling.}\label{tab:relaydestbefore}
	{\footnotesize
	\begin{tabular}{|c|c|c|c|c|c|c|}
		\hline
		\backslashbox{Symbol}{Time} & $t$         & $t+1$           & $t+2$                       & $t+3$                                         & $t+4$                                           & $t+5$                                           \\ \hline
		$\bx'[1]$                                      & $\bs_{t-2}[3]$ & $\bs_{t-1}[3]$ & $\bs_{t}[3]$             & $\bs_{t+1}[3]$                               & $\bs_{t+2}[3]$                                 & $\bs_{t+3}[3]$                                 \\ \hline
		$\bx'[2]$                                      & $\bs_{t-3}[2]$ & $\bs_{t-2}[2]$ & $\bs_{t-1}[2]$             & $\bs_{t}[2]$                               & $\bs_{t+1}[2]$                                 & $\bs_{t+2}[2]$                                 \\ \hline
		$\bx'[3]$                                      & $\bs_{t-4}[1]$ & $\bs_{t-3}[1]$ & $\bs_{t-2}[1]$             & $\bs_{t-1}[1]$                               & $\bs_{t}[1]$                                 & $\bs_{t+1}[1]$                                 \\ \hline
		$\bx'[4]$                                      &             & $\bs_{t-4}[1]$     &
		\begin{tabular}{c}$\bs_{t-3}[1]$ \\ $+\bs_{t-3}[2]$\end{tabular} & 
		\begin{tabular}{c}$\bs_{t-2}[1] + \bs_{t-2}[2]$ \\ $+\bs_{t-2}[3]$\end{tabular} & \begin{tabular}{c}$\bs_{t-1}[1] + \bs_{t-1}[2]$ \\ $+\bs_{t-1}[3]$\end{tabular} & 
		\begin{tabular}{c}$\bs_{t}[1] + \bs_{t}[2]$ \\ $+\bs_{t}[3]$\end{tabular}
		\\ \hline
	\end{tabular}
	}
	\end{subtable}
	
	\begin{subtable}[]{\textwidth}
	\caption{Diagonal Interleaving $(4, 3)$ MDS code after relabeling.}\label{tab:relaydestafter}
	{\footnotesize
	\begin{tabular}{|c|c|c|c|c|c|c|}
		\hline
		\backslashbox{Symbol}{Time} & $t$         & $t+1$           & $t+2$                       & $t+3$                                         & $t+4$                                           & $t+5$                                           \\ \hline
		$\bx'[1]$                                      & $\bs'_t[1]$ & $\bs'_{t+1}[1]$ & $\bs'_{t+2}[1]$             & $\bs'_{t+3}[1]$                               & $\bs'_{t+4}[1]$                                 & $\bs'_{t+5}[1]$                                 \\ \hline
		$\bx'[2]$                                      & $\bs'_t[2]$ & $\bs'_{t+1}[2]$ & $\bs'_{t+2}[2]$             & $\bs'_{t+3}[2]$                               & $\bs'_{t+4}[2]$                                 & $\bs'_{t+5}[2]$                                 \\ \hline
		$\bx'[3]$                                      & $\bs'_t[3]$ & $\bs'_{t+1}[3]$ & $\bs'_{t+2}[3]$             & $\bs'_{t+3}[3]$                               & $\bs'_{t+4}[3]$                                 & $\bs'_{t+5}[3]$                                 \\ \hline
		$\bx'[4]$                                      &             & $\bs'_t[3]$     &
		\begin{tabular}{c}$\bs'_{t+1}[3]$ \\ $+\bs'_t[2]$\end{tabular} & 
		\begin{tabular}{c}$\bs'_{t+2}[3] + \bs'_{t+1}[2]$ \\ $+\bs'_{t}[1]$\end{tabular} & \begin{tabular}{c}$\bs'_{t+3}[3] + \bs'_{t+2}[2]$ \\ $+\bs'_{t+1}[1]$\end{tabular} & 
		\begin{tabular}{c}$\bs'_{t+4}[3] + \bs'_{t+3}[2]$ \\ $+\bs'_{t+2}[1]$\end{tabular}
		\\ \hline
	\end{tabular}
	}
	\end{subtable}
\end{table*}

Further, note that, with respect to $\bs'$, the relay node transmits using a systematic code, although such code is not systematic with respect to the source packets $\bs$.

\subsection{Example~2}
\label{ex:concatenation}
	Consider an $(3, 2, [1, 2])_{\FF}$ and an $(4, 3, [1, 2, 3])_{\FF}$ point-to-point single-link codes under 1 erasure. The achievable delay spectrum are $\bT' = [2, 1]$ and $\bT'' = [3, 2, 1]$, respectively. The alternate description is given by $\bG' = [(2, 1), (1, 1)]$ and $\bG'' = [(3,1), (2, 1), (1, 1)]$, that is, there is one symbol for each delay in each code. Then, the concatenation of both can be described with $$\bG = [ (3, 0+1), (2, 1+1), (1, 1+1)] = [(3, 1), (2, 2), (1, 2)].$$ Such concatenation is shown in Table~\ref{tab:exampcode}.
	
	\begin{table}
	    \centering
	    \caption{Example of $(7, 5, [1, 2, 1, 2, 3])_{\FF}$ point-to-point single-link code resulting from the concatenation of an $(3, 2, [1, 2])_{\FF}$ and an $(4, 3, [1, 2, 3])_{\FF}$ codes.}\label{tab:exampcode}
        \begin{tabular}{|c|c|c|c|c|}
        \hline
        \backslashbox{Symbol}{Time} & $t$          & $t+1$          & $t+2$   & $t+3$                     \\ \hline
        $\bx[1]$                                       & $\bs_{t}[1]$ & $\bs_{t+1}[1]$ & $\bs_{t+2}[1]$  & $\bs_{t+3}[1]$              \\ \hline
        $\bx[2]$                                       & $\bs_{t}[2]$ & $\bs_{t+1}[2]$ & $\bs_{t+2}[2]$         & $\bs_{t+3}[2]$      \\ \hline
        $\bx[3]$                                       &              & $\bs_{t}[1]$   & $\bs_{t+1}[2] + \bs_{t}[1] $ & $\bs_{t+2}[2] + \bs_{t + 1}[1]$ \\ \hline
        $\bx[4]$                                       & $\bs_{t}[3]$ & $\bs_{t+1}[3]$ & $\bs_{t+2}[3]$     & $\bs_{t+3}[3]$          \\ \hline
        $\bx[5]$                                       & $\bs_{t}[4]$ & $\bs_{t+1}[4]$ & $\bs_{t+2}[4]$     & $\bs_{t+3}[4]$          \\ \hline
        $\bx[6]$                                       & $\bs_{t}[5]$ & $\bs_{t+1}[5]$ & $\bs_{t+2}[5]$     & $\bs_{t+3}[5]$          \\ \hline
        $\bx[7]$                                       &              & $\bs_{t}[5]$   & $\bs_{t+1}[5] + \bs_{t}[4]$ & $\bs_{t+2}[5] + \bs_{t+1}[4] + \bs_t[3]$ \\ \hline
        \end{tabular}
    \end{table}
	
	Note that, since the possible delays are different, the operation is not simply summing $\bk'^{(g)} + \bk''^{(g)}$ (which can not be done since both have different lengths), rather, the correct delays must be considered. Alternatively, we can describe all relevant codes under a common possible delay spectrum, in that case, we would have $\bG' = [(3, 0), (2, 1), (1, 1)]$, and the resulting concatenation can be described simply by the sum.
\subsection{Example~3}
\label{ex:firststep}
    As an example, let us compute, step by step, the achieved rate for the setting in Fig.~\ref{fig:1t2ex}. We start by computing the upper bounds which are shown in the figure, and notice that the first hop is the bottleneck. We set $n = 12$, $k^{(1)} = 8$, and compute the delay spectrum $\bG^{(1)}$ and the constraint $\bG^{\textrm{con}}$ which are
    \begin{align*}
        \bG^{(1)} =\begin{bmatrix} (2, 4) & (1, 4) \end{bmatrix}~;~\bG^{\textrm{con}} = \begin{bmatrix} (3, 4) & (2, 4) & (1, 0)\end{bmatrix}.
    \end{align*}
    This constraint means we can transmit, at most, 4 symbols with delay 3, and 4 symbols with delay 2.
    
    Then, we move on to the first link in the second hop, that is, the one with 3 erasures (remember we start with the link with most erasures). We have $\bT_1 = [3, 2, 1]$ and $\bR^{\textrm{con}} = [0, 1/3, 2/3]$. We then compute
    \begin{align*}
        \bk'[1] = 12 - 12 \cdot 3 \left( \frac{1}{4} \right) = 3,~\bk'[2] = 12 - 12\cdot 3 \left( \frac{2/3}{3}\right) = 4
    \end{align*}
    and set $k^{(2)}_1 = 3$. We then compute the delay spectrum for this link, which is given by 3 symbols with delay 3, update the constraint to
    \begin{equation*}
        \bG^{\textrm{con}} = \begin{bmatrix} (3, 1) & (2, 4) & (1, 0) \end{bmatrix}
    \end{equation*}
    and move on to the second link. Note that $\bT_i$ is the same, but the constraint $\bR^{\textrm{con}} = [0, 1/12, 5/12]$ has been loosened due to the symbols already transmitted. Now, we again compute
    \begin{align*}
        \bk'[1] = 12 - 12 \cdot 2 \left( \frac{1}{4} \right) = 6,~\bk'[2] = 12 - 12 \cdot 2 \left( \frac{11/12}{3} \right) = 4.666,~\bk'[3] = 12 - 12 \cdot 2 \left( \frac{7/12}{2} \right) = 5
    \end{align*}
    and set $k^{(2)}_2 = 4$. Note that, in this case, this code is suboptimal, i.e., we are not able to use 4.666. This could be fixed by choosing $n = 36$, but we will improve this rate in the following section. Computing the delay spectrum, we get 4 symbols transmitted with delay 2.
    
    Note that we have $k^{(1)} = 8$ and $k^{(2)} = 7$, thus, we must erase one of the symbols in the first hop, i.e., it can carry no information. This is improved in Section~\ref{sec:optimizationbn}.

\subsection{Example~4}
\label{ex:algorithm}
	We start by computing $R^{(1)} = 1$ and $R^{(2)} = 5/4$, which was done previously. In this case, $R^{(1)}$ is the bottleneck. We set
	\begin{align*}
	n = 20, \quad k_1^{(1)} = 12, \quad k_2^{(1)} = 8
	\end{align*}
	and compute the delay spectra for these codes. This will result in the following equally-delayed groups:
	\begin{align}
	\bG^{(1)}_1 = \begin{bmatrix}
	(4, 4) & (3, 4) & (2, 4)
	\end{bmatrix}, \quad 
	\bG^{(1)}_2 = \begin{bmatrix}
	(4, 4) & (3, 4) & (2, 0)
	\end{bmatrix}
	\end{align}
	
	The way to achieve these delay spectra is described in the example in Section~\ref{sec:sepcode}, and these delay spectra are generally achievable as shown in Lemma~\ref{lemma:achiev}.
	
	Then, we can compute
	\begin{align}
	\bk^{(1)} = [8, 8, 4]
	\end{align}
	and the tuple constraint
	\begin{align}
	\bG^{\textrm{con}} = \begin{bmatrix}
	(3, 4) & (2, 8) & (1, 8) & (0, 0)
	\end{bmatrix}
	\end{align}
	thus $\bk^{\textrm{con}} = [4, 8, 8, 0]$. 
	
	For the remaining of the example, recall that the links are ordered in decreasing order of number of erasures, that is, $N^{(2)}_1 = 2$ and $N^{(2)}_2 = 1$, which is inverted w.r.t. the figure. 
	
	Now, for the second hop, we have
	\begin{align*}
	\bT_1 = [3, 2, 1, 0], \quad 
	\bR^{\textrm{con}} = [0, \frac{4}{20}, \frac{12}{20}, \frac{20}{20} ]
	\end{align*}
	and we solve
	\begin{align*}
		\bk'[1] &= 20 - 20 \cdot 2 \left( \frac{1 - 0}{3 + 1}  \right) = 10 , \quad 
		\bk'[2] = 20 - 20 \cdot 2 \left( \frac{1 - 4/20}{2 + 1}  \right) = 9.33 \\
		\bk'[3] &= 20 - 20 \cdot 2 \left( \frac{1 - 12/20}{1 + 1}  \right) = 12
	\end{align*}
	and we set $k^{(2)}_1 = 9$. Now, note that $20 - 9 = 11$ is not a multiple of $N^{(2)}_1 = 2$, thus, we need to adjust the parameters in order to make it a multiple. Thus, we update $n = 40$, $k^{(1)}_1 = 24$, $k^{(1)}_2 = 16$ and $k^{(2)}_1 = 18$. The code used in the first hop is constructed simply by concatenating the same codes used previously twice. Then, we also update the constraint, i.e., we would have
	\begin{align}
	\bG^{\textrm{con}} = \begin{bmatrix}
	(3, 8) & (2, 16) & (1, 16) & (0, 0)
	\end{bmatrix}
	\end{align}
	and $\bk^{\textrm{con}} = [8, 16, 16, 0]$. 
	
	Now, we compute the delay spectra for the first link of the second hop. Using our achievability result, we get
	\begin{align}
	\bG^{(2)}_1 = \begin{bmatrix}
	(3, 7) & (2, 11) & (1, 0) & (0, 0)
	\end{bmatrix}
	\end{align}
	that is, we transmit 11 symbols with delay 2 and 7 symbols with delay 3. This is achievable using a concatenation of seven $(4, 2)$ and four $(3, 1)$ diagonal interleaving MDS codes. Notice that this is slight different than our example due to the constraint in $n$ being equal to all links.
	
	Then, we need to update $\bk^{\textrm{con}}$. Note that, by making the delay structure the same for the constraint and for the code, this can be done by simply subtracting the second field of each tuple, i.e., $\bk^{\textrm{con}} = [8 - 7, 16 - 11, 16 - 0, 0 - 0] = [1, 5, 16, 0]$. 	
	
	We now proceed to solve for $i = 2$. Again, we have
	\begin{align*}
	\bT_2 = [3, 2, 1, 0], \quad \bR^{\textrm{con}} = [0, \frac{1}{40}, \frac{6}{40}, \frac{22}{40}].
	\end{align*}
	We solve
	\begin{align*}
	\bk'[1] &= 40 - 40 \cdot 1 \left( \frac{1 - 0 }{3 + 1} \right) = 30, \quad
	\bk'[2] = 40 - 40 \cdot 1 \left( \frac{1 - 1/40 }{2 + 1} \right) = 27 \\ 
	\bk'[3] &= 40 - 40 \cdot 1 \left( \frac{1 - 6/40}{1 + 1} \right) = 23, \quad
	\bk'[4] = 40 - 40 \cdot 1 \left( \frac{1 - 22/40}{0 + 1} \right) = 22.
	\end{align*}
	Finally, we set $k^{(2)}_2 = 22$ and find the delay spectrum, which is
	\begin{equation*}
	\bG^{(2)}_2 = \begin{bmatrix}
	(3, 0) & (2, 4) & (1, 18) & (0, 0)
	\end{bmatrix}.
	\end{equation*}
	Note that, in this case, updating $\bk^{\textrm{con}}$ would result in $[1 - 0, 5 - 4, 16 - 18, 0 - 0] = [1, 1, -2, 0]$. We would then update it to $[0,0,0,0]$. Essentially, it means that two symbols will arrive earlier at the destination than the deadline, i.e., by times 3 and 4, instead of 5.
	
	On the other hand, the baseline scheme achieves a rate of 0.75. This is achieved with $T_{mw}^{(1)} = 3$ and $T_{mw}^{(2)} = 2$. Thus, even this simple example shows that there is a considerable improvement that can be made using our proposed rates rather than the simple baseline scheme.

\end{document}